\documentclass[aps,prl,superscriptaddress,amsmath,amssymb,twocolumn,showpacs,floatfix,english,noshowpacs,longbibliography]{revtex4-1}

\usepackage{url}
\usepackage{graphicx}
\usepackage{color}
\usepackage[colorlinks=true, urlcolor=blue, linkcolor=blue, citecolor=blue, pdftex]{hyperref}
\usepackage[english]{babel}
\usepackage{braket}
\usepackage{amssymb}
\usepackage{subfigure}

\DeclareGraphicsExtensions{.pdf,.png,.jpg}

\newcommand{\ham}{\hat{\mathcal{H}}}
\newcommand{\Sh}{\hat{\textbf{S}}}
\newcommand{\up}{\uparrow}
\newcommand{\down}{\downarrow}

\AtBeginDocument{%
    \newwrite\bibnotes
    \def\bibnotesext{Notes.bib}
    \immediate\openout\bibnotes=\jobname\bibnotesext
    \immediate\write\bibnotes{@CONTROL{REVTEX41Control}}
    \immediate\write\bibnotes{@CONTROL{%
    apsrev41Control,author="08",editor="1",pages="1",title="0",year="1"}}
     \if@filesw
     \immediate\write\@auxout{\string\citation{apsrev41Control}}%
    \fi
}

\begin{document}

\title{Triplons, Magnons, and Spinons in a Single Quantum Spin System: 
SeCuO$_3$}

\author{Luc Testa}
\email[]{luc.testa@gmail.com}
\affiliation{Institute of Physics, Ecole 
Polytechnique F\'ed\'erale de Lausanne (EPFL), CH-1015 Lausanne, Switzerland}

\author{Vinko \v{S}urija }
\affiliation{Institute of Physics, Bijeni\v{c}ka 46, HR-10000 Zagreb, Croatia}

\author{Krunoslav Pr\v{s}a }
\affiliation{Institute of Physics, Ecole 
Polytechnique F\'ed\'erale de Lausanne (EPFL), CH-1015 Lausanne, Switzerland}

\author{Paul Steffens }
\affiliation{Institut Laue-Langevin, BP 156, 38042 Grenoble Cedex 9, France}

\author{Martin Boehm }
\affiliation{Institut Laue-Langevin, BP 156, 38042 Grenoble Cedex 9, France}

\author{Philippe Bourges }
\affiliation{Universit\'e Paris-Saclay, CNRS, CEA, Laboratoire L\'eon 
Brillouin, 91191, Gif-sur-Yvette, France}

\author{Helmuth Berger}
\affiliation{Institute of Physics, Ecole 
Polytechnique F\'ed\'erale de Lausanne (EPFL), CH-1015 Lausanne, Switzerland}

\author{Bruce Normand}
\affiliation{Paul Scherrer Institute, CH-5232 Villigen PSI, Switzerland}
\affiliation{Lehrstuhl f\"ur Theoretische Physik I, Technische Universit\"at 
Dortmund, Otto-Hahn-Strasse 4, 44221 Dortmund, Germany}
\affiliation{Institute of Physics, Ecole 
Polytechnique F\'ed\'erale de Lausanne (EPFL), CH-1015 Lausanne, Switzerland}

\author{Henrik M. R\o nnow  }
\email[]{henrik.ronnow@epfl.ch}
\affiliation{Institute of Physics, Ecole 
Polytechnique F\'ed\'erale de Lausanne (EPFL), CH-1015 Lausanne, Switzerland}

\author{Ivica \v{Z}ivkovi\'c  }
\affiliation{Institute of Physics, Ecole 
Polytechnique F\'ed\'erale de Lausanne (EPFL), CH-1015 Lausanne, Switzerland}

\date{\today}

\begin{abstract}
Quantum magnets display a wide variety of collective excitations, including 
spin waves (magnons), coherent singlet-triplet excitations (triplons), and 
pairs of fractional spins (spinons). These modes differ radically in nature 
and properties, and in all conventional analyses any given material is 
interpreted in terms of only one type. We report inelastic neutron scattering 
measurements on the spin-1/2 antiferromagnet SeCuO$_3$, which demonstrate that 
this compound exhibits all three primary types of spin excitation. Cu$_1$ sites 
form strongly bound dimers while Cu$_2$ sites form a network of spin chains, 
whose weak three-dimensional (3D) coupling induces antiferromagnetic order. 
We perform quantitative modeling to extract all of the relevant magnetic 
interactions and show that magnons of the Cu$_2$ system give a lower 
bound to the spinon continua, while the Cu$_1$ system hosts a band of 
high-energy triplons at the same time as frustrating the 3D network. 
\end{abstract}

\maketitle

The exotic collective excitations observed in magnetic materials emerge from 
the rich spectrum of possible effects when quantum spin fluctuations act in 
different geometries, dimensionalities, and with different degrees of 
frustration. When fluctuations push a system beyond robust magnetic order 
and textbook spin waves, common types of excitation include triplons arising 
from structural dimerization \cite{Cavadini2001,Thielemann2009,Schmidiger2013} 
and frustration \cite{Gaulin2004}, bound states of magnons and triplons 
\cite{Torrance1969,Uhrig1996,Ward2017,McClarty2017}, and fractions of magnons 
and triplons; these latter include spinons \cite{Hammar1999,Thielemann2009,
Mourigal2013}, solitons \cite{Nagler1983,Faure2018}, Majorana quasiparticles 
\cite{Kitaev2006,Chaloupka2010,Wang2019}, and other topological objects. In 
all these situations, the system is normally analyzed in terms of just one 
type of excitation, and detailed theoretical and numerical approaches have 
been developed for comparison with experiment.

However, an often overlooked category is the set of quantum magnets in which 
magnetic order is present only as a rather thin veneer on a ``background'' 
dominated by quantum fluctuations. While the Bragg peaks and magnons of the 
ordered spin component tend to dominate the measured experimental response, 
no rule states that the remaining spin fluctuations should be incoherent. 
The field- \cite{Rueegg2003} and pressure-induced \cite{Rueegg2008} quantum 
phase transitions of TlCuCl$_3$ provide an example of arbitrarily weak 
antiferromagnetic order superposed on a fluctuating dimer system with 
triplon excitations. In KCuF$_3$, weak coupling of the spin chains produces 
magnetic order superposed on a system of spinons, which is revealed at high 
energy \cite{Lake2005}. Candidate spinon excitations also coexist with 
square-lattice antiferromagnetism \cite{DallaPiazza2015}, and many 
low-dimensional metal-organic systems offer the possibility of controling 
this coexistence \cite{Skoulatos2017}. One structural route to the same 
phenomena is provided by the spin-tetrahedron material Cu$_2$Te$_2$O$_5$X$_2$ 
(X = Cl, Br) \cite{Johnsson2000}, where the magnetic response is dominated by 
the non-ordered spins \cite{Jaglicic2006}, a situation anticipated in theory 
\cite{Kotov2004}, and incommensurate magnetism \cite{Zaharko2004} coexists 
with anomalous coupled-cluster excitations \cite{Prsa2009}; similar 
coupled-cluster physics has been pursued in a number of other materials 
\cite{Becker2005,Zaharko2008,Choi2014}. Another is the recent discovery 
\cite{Zhang2020} of both spinons and magnons in very weakly coupled chains 
of two different types.

\begin{figure*}[t]
\centering
\includegraphics[width=\textwidth]{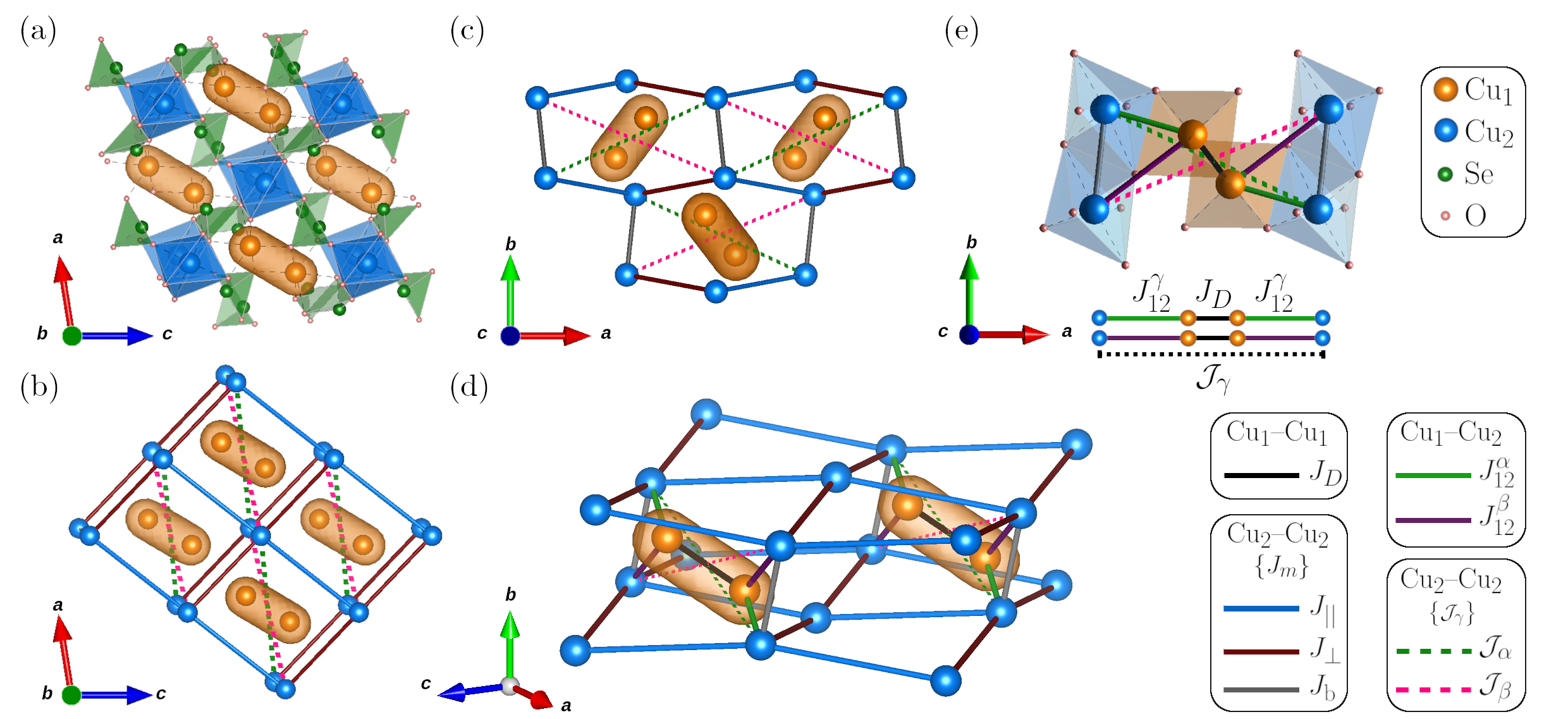}
\caption{{\bf Atomic structure and magnetic interactions of SeCuO$_3$.} 
(a) Schematic representation of the atomic structure showing Cu$_1$ 
(orange), Cu$_2$ (blue), Se (green), and O (pink) atoms. (b) Projection on 
the $ac$ and (c) on the $ab$ plane, indicating the magnetic interactions 
of Table \ref{tab_Jparam}. (d) Perspective view highlighting the Cu$_2$ 
chains and the direct interactions connecting them into coupled, buckled 
planes. (e) Geometry of the effective interactions mediated between Cu$_2$ 
atoms by the Cu$_1$ dimer units: $\mathcal{J}_{\alpha}$ and $\mathcal{J}_{\beta}$ 
are given in terms of the two different Cu$_1$--Cu$_2$ interactions, 
$J_{12}^\alpha$ (green) and $J_{12}^\beta$ (purple), and the dimer interaction, 
$J_D$ (black), by Eq.~(\ref{eqt_tetramer_jeff}).}
\label{fig:structure}
\end{figure*}

Here we investigate the coexistence of multiple excitation types by 
an inelastic neutron scattering (INS) study of SeCuO$_3$. This compound 
displays both quasi-localized high-energy states and weak magnetic order 
at low temperatures. We demonstrate that the excitation spectrum has one 
triplon branch, dispersing weakly around 27 meV, and a magnon-like branch 
below 4 meV, whose associated scattering intensity shows the clear 
fingerprints of spinon continua. By model calculations using linear spin-wave 
theory and perturbative methods, we deduce the interaction parameters of a 
minimal magnetic Hamiltonian, allowing us to describe SeCuO$_3$ in terms of 
two interacting spin subsystems, namely dimers and chains, each of which 
shapes the magnetic excitations of the other. 

The $S = 1/2$ quantum magnet SeCuO$_3$ \cite{kristall113} has a monoclinic 
unit cell with space group P2$_1$/n and lattice parameters $a = 7.71$ \AA, 
$b = 8.24$ \AA, $c = 8.50$ \AA, and $\beta = 99.12^{\circ}$. Two 
crystallographically inequivalent Cu sites, Cu$_1$ and Cu$_2$, are each 
surrounded by six O atoms, forming CuO$_4$ plaquettes, with the remaining two 
O atoms forming the elongated octahedra represented for the Cu$_2$ atoms in 
Figs.~\ref{fig:structure}(a) and \ref{fig:structure}(e). This elongation 
favors the $d_{x^2-y^2}$ orbitals, ensuring strong Cu$_1$ dimer units (orange 
shading in Fig.~\ref{fig:structure}) of edge-sharing plaquettes whose 
superexchange paths have a Cu$_1$--O--Cu$_1$ angle of 101.9$^\circ$ 
\cite{SeCuO3_2012_bulk}. Including the Cu$_2$ $d_{x^2-y^2}$ orbitals 
led to the proposal of a weakly coupled network of linear 
(Cu$_2$--Cu$_1$--Cu$_1$--Cu$_2$) tetramers \cite{SeCuO3_2012_bulk}, although 
this scenario cannot explain the magnetic susceptibility below 90 K. Recent 
nuclear quadrupole resonance (NQR) measurements confirmed the formation of 
singlet states at $T \lesssim 200$ K \cite{SeCuO3_diffraction}, and together 
with nuclear magnetic resonance (NMR), electron spin resonance (ESR), and 
torque magnetometry experiments \cite{secuo3_lee_dec,SeCuO3_decoupled} were 
interpreted as reflecting two essentially decoupled subsystems, the strong, 
local Cu$_1$ dimers and weakly coupled Cu$_2$ spins hosting magnetic order 
below $T_N = 8$ K.

To access the full spin dynamics of SeCuO$_3$, we grew a 1 g single-crystal 
sample by chemical vapor transport. Thermal neutron INS measurements 
were performed on the IN8 spectrometer (ILL \cite{DOI_IN8}) to probe the 
(\textit{hkh}) scattering plane. The low-energy dynamics were studied on the 
cold-neutron spectrometers ThALES (ILL \cite{DOI_thales}) and 4F1 (LLB), the 
latter experiment probing the (\textit{hk$\bar{h}$}) scattering plane. Full 
details of the instrumental set-ups employed are provided in Sec.~S1 of the 
SM \cite{sm}. The measured intensities, $I({\bf q},\omega)$, are directly 
proportional to the dynamical structure factor, $S({\bf q},\omega)$, for 
scattering processes at momentum transfer ${\bf q}$ and energy transfer 
$\omega$.

\begin{figure}[t]
\centering
\includegraphics[width=\linewidth]{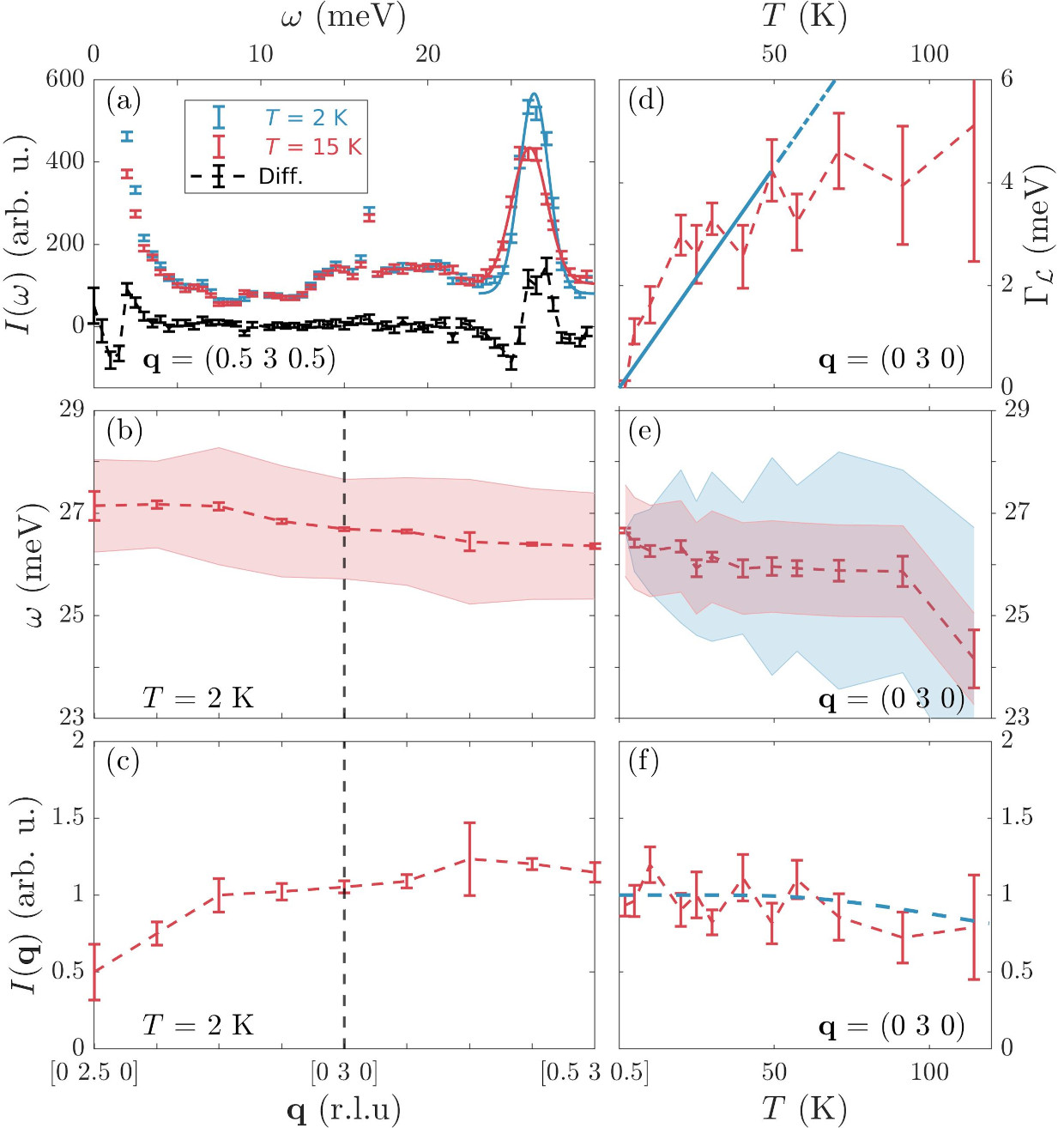}
\caption{{\bf Triplon excitation.} (a) Intensity, $I(\omega)$, at $\textbf{q}
 = (0.5 \; 3 \; 0.5)$, measured at low (blue) and intermediate (red) 
temperatures with the difference shown in black. (b) Dispersion, $\omega 
(\textbf{q})$, of the triplon in two orthogonal $\textbf{q}$ directions; 
shading represents the width (FWHM) of the mode at each \textbf{q} point. 
(c) Integrated intensity, $I({\bf q})$, for the same directions. (d)-(f) 
Thermal evolution at $\textbf{q} = (0 \; 3 \; 0)$. (d) Lorentzian width, 
$\Gamma_{\mathcal{L}}$ (red), compared to $k_{\rm B} T$ (blue). (e) $\omega 
(\textbf{q},T)$; shading indicates the instrumental resolution of 1.8(2) 
meV (red) and the Lorentzian profile (blue). (f) Normalized $I({\bf q})$ 
in red compared with the thermal singlet population (blue).}
\label{highE}
\end{figure}

In Fig.~\ref{highE} we present the high-energy dynamics of SeCuO$_3$ as 
measured on IN8. We obtained $I({\bf q},\omega)$ for \textbf{q} points 
along two orthogonal high-symmetry directions. At 2 K, each energy scan 
[Fig.~\ref{highE}(a)] has a resolution-limited peak that we fit with a 
Gaussian at all ${\bf q}$ points to extract a weak dispersion around 26.5 
meV [Fig.~\ref{highE}(b)], with smooth changes in intensity 
[Fig.~\ref{highE}(c)]. At 15 K, i.e.~above $T_N$, the peak shows only a 
minimal downward shift and increased broadening [Fig.~\ref{highE}(a)]. 
Figures \ref{highE}(d)-\ref{highE}(f) confirm that this mode persists at 
least until 120 K, i.e.~far beyond $T_N$, and that its width is captured 
by the Lorentzian component of a Voigt line shape. 

The weak ${\bf q}$-dependence of this excitation indicates its nature as a 
near-localized triplon of the Cu$_1$ dimers, whose energy is given by $J_D$ 
in Fig.~\ref{fig:structure}. Its Lorentzian width increases linearly with 
temperature until a value of 4 meV [Figs.~\ref{highE}(d) and \ref{highE}(e)], 
which we will show reflects the coupling to the excitations of the Cu$_2$ 
subsystem. However, the primary thermal effect is intrinsic, as shown in 
Fig.~\ref{highE}(f) by comparing the mode amplitude with the probability, 
$[1 + 3 \exp(- J_D/k_\text{B} T)]^{-1}$, of finding a $J_D$ dimer in its 
singlet state at temperature $T$, which further confirms the triplon 
nature of this mode.

\begin{figure}[t]
\centering
\includegraphics[width=\linewidth]{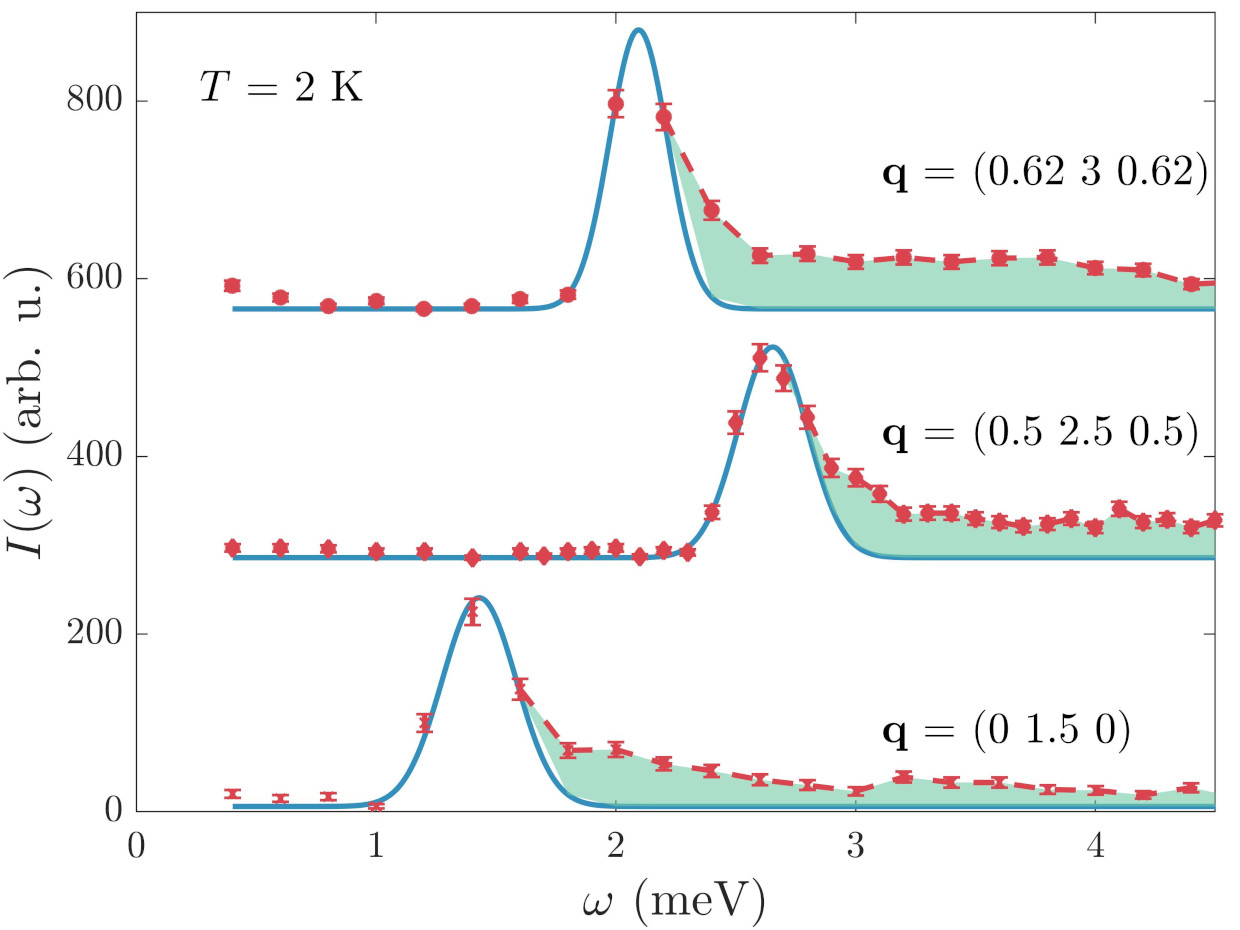}
\vspace{-5mm}
\caption{{\bf Magnon peak and scattering continua.} $I(\omega)$ at three 
different $\textbf{q}$ points. Measured data (red points) are fitted by a 
Gaussian peak (blue line) at the lower edge. Green shading indicates a 
scattering continuum at higher energies.} 
\label{lowE_cuts}
\end{figure}

\begin{figure*}[t]
\centering
\includegraphics[width=\linewidth]{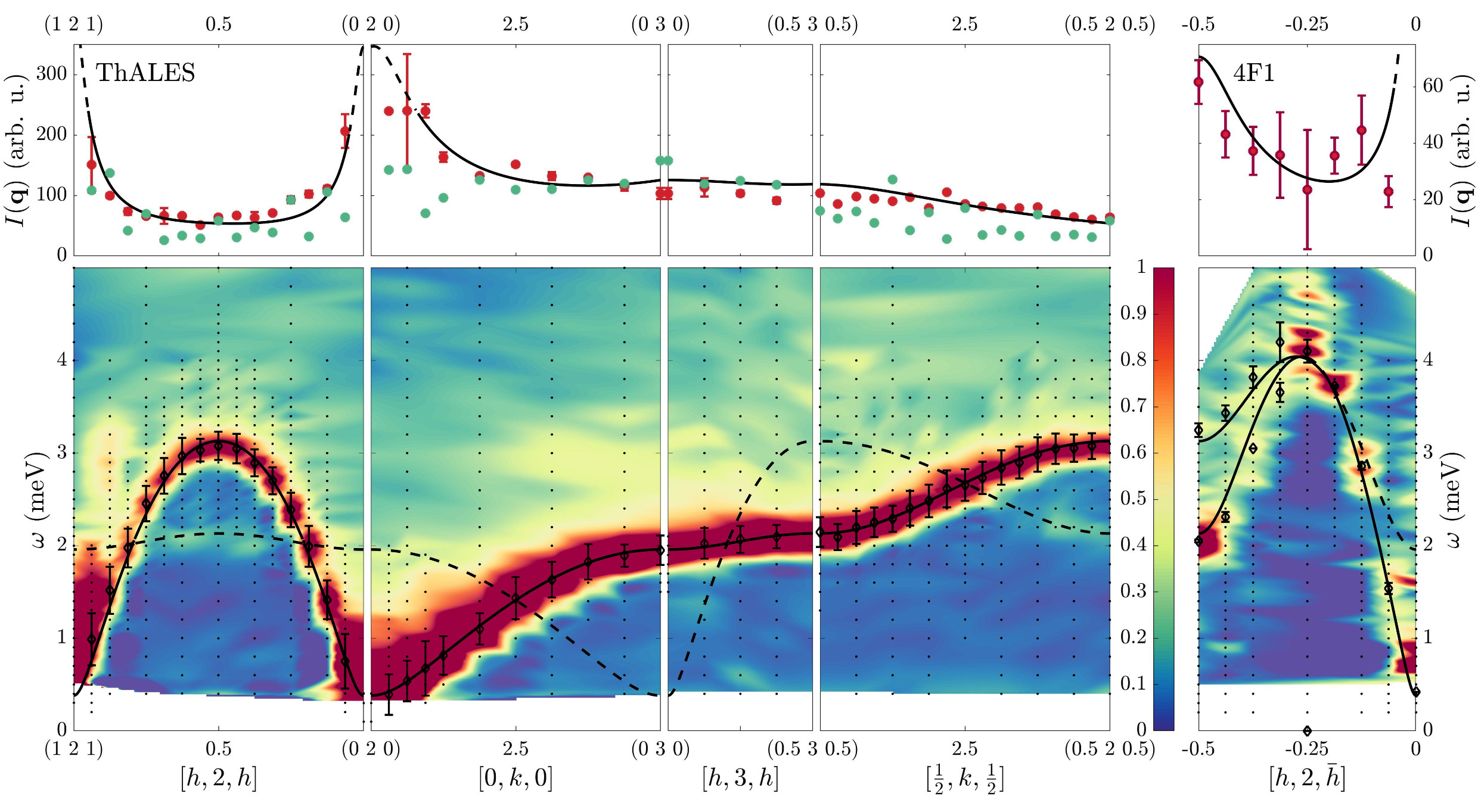}
\caption{{\bf Magnon and spinon spectra.} Colored panels show the scattering 
intensity, $S(\textbf{q},\omega)$, for five different $\textbf{q}$ directions. 
The black points show the peak centers taken from Fig.~\ref{lowE_cuts} and 
the error bars the extracted width (FWHM) of the Gaussian profiles. The black 
lines show the two spin-wave branches in the model fit to the lower peak 
of Fig.~\ref{lowE_cuts}, one of which (dashed line) has vanishing intensity. 
The upper panels show the integrated intensities, $I({\bf q})$, of the measured 
spin-wave contribution [$I_{\rm p}({\bf q})$, red points] and the continuum 
contribution [$I_{\rm c}({\bf q})$, green points], taken respectively from the 
peak and shaded areas in Fig.~\ref{lowE_cuts}; the black lines show the 
spin-wave intensities given by the model parameters. In the panel at right, 
the intensity is a sum of two modes.}
\label{swfitandintensities}
\end{figure*}

Turning to the low-energy dynamics measured at 2 K on ThALES and 4F1, 
representative background-subtracted constant-${\bf q}$ $\omega$ scans are 
shown in Fig.~\ref{lowE_cuts}. A strong low-energy mode is present at all 
${\bf q}$, but a continuum of scattering states persists above this feature, 
at least up to the highest measured energy of 4.5 meV. To visualize this 
continuum scattering, we present our intensity data as color maps in 
Fig.~\ref{swfitandintensities}, and note that it appears in all three 
dimensions of reciprocal space. We return below to a detailed discussion 
of this continuum. 

For a systematic analysis we perform a Gaussian fit to the peak at the 
lower edge of the continuum (Fig.~\ref{lowE_cuts}) and collect two separate 
intensity contributions, $I_{\rm p}$ from the Gaussian and $I_{\rm c}$ from the 
excess scattering at all higher energies. The upper panels of 
Fig.~\ref{swfitandintensities} show the values of $I_{\rm p}(\textbf{q})$ 
and $I_{\rm c}(\textbf{q})$ extracted from 74 energy scans. The lower panels 
show a well-defined band with a maximum of 4 meV and a small gap, $\Delta = 
0.42(3)$ meV, where $I_{\rm p}(\textbf{q})$ becomes large due to the magnetic 
order.

We expect that, with the exception of a term opening the gap, the minimal 
magnetic Hamiltonian contains only Heisenberg interactions between 
near-neighbor spins in all directions, and thus takes the form  
\begin{equation}\label{eq_hamiltonian} 
\ham =  J_D \!\!\! \sum_{\braket{i_1,j_1}} \!\!\! \Sh_{i_1} \! \cdot \Sh_{j_1} \! + 
\!\!\!\! \sum_{\braket{i_1,i_2},\gamma} \!\!\!\!\! J_{12}^\gamma \Sh_{i_1} \! \cdot 
\Sh_{i_2} + \!\!\!\! \sum_{\left\lbrack i_2,j_2 \right\rbrack_{m}} \!\!\!\! J_{m} \, 
\Sh_{i_2} \! \cdot \Sh_{j_2}. 
\end{equation}
Here $i_1$ and $j_1$ denote Cu$_1$ sites and $i_2$ and $j_2$ Cu$_2$ sites, 
$\langle \dots \rangle$ denotes a sum restricted to nearest-neighbor bonds 
and $[ \dots ]_m$ a sum over bonds in the set $\{J_m\}$. Having interpreted 
the high-energy response as a triplon of the Cu$_1$ subsystem with $J_D = 26.5$ 
meV, we build up our knowledge of the terms in Eq.~(\ref{eq_hamiltonian}) by 
next describing the low-energy response as a consequence of the decoupled 
Cu$_2$ subsystem, i.e.~by neglecting the second term.
For this we seek a set of interaction parameters that, used in an effective 
Hamiltonian of the same form as the last term of Eq.~(\ref{eq_hamiltonian}), 
reproduces the magnon dispersions and intensities in 
Fig.~\ref{swfitandintensities}. As shown in Fig.~\ref{fig:structure}, we 
allow both near-neighbor couplings, $\{J_m\}$, and long-distance ``effective'' 
couplings, $\{\mathcal{J}_\gamma\}$, over paths that include the polyhedra of 
other Cu$_1$ atoms and whose microscopic origin therefore lies in the second 
term of Eq.~(\ref{eq_hamiltonian}), as we discuss further below. 

We fit $\omega (\textbf{q})$ using linear spin-wave (LSW) theory, as 
implemented in the SpinW package \cite{spinw}, obtaining an excellent 
account of the INS peak positions when the interaction parameters of 
Fig.~\ref{fig:structure} have the values reported in Table \ref{tab_Jparam}. 
This fit contains two 
magnon branches, one of which has over 90\% of the intensity we measure. 
The LSW theory delivers an accurate account of $I_{\rm p} (\textbf{q})$ for 
the strong branch with no further fitting, as Fig.~\ref{swfitandintensities} 
makes clear, underlining its success in capturing all the leading physics of 
the magnon spectrum. Turning to percent-level discrepancies, in the LSW 
treatment the weak branch has a very low [$\mathcal{O}$(0.1\%)] intensity, 
whereas the intensities measured in Fig.~\ref{swfitandintensities} are in 
general 1-5\% of the strong branch. While this discrepancy suggests that the 
magnetic Hamiltonian of SeCuO$_3$ contains further terms (beyond those in 
Table \ref{tab_Jparam}) coupling two quasi-independent magnetic sublattices, 
the low intensity indicates that they are very weak.

\begin{table}[b]
\caption{\label{tab_Jparam}
Superexchange interaction parameters within the Cu$_1$ and Cu$_2$ subsystems, 
in meV, obtained by fitting the dispersion data of Figs.~\ref{highE}(b) and 
\ref{swfitandintensities}. The geometry of the interactions is shown in 
Fig.~\ref{fig:structure}.}
\begin{ruledtabular}
\begin{tabular}{c|ccccc}
Cu$_1$--Cu$_1$ & & & Cu$_2$--Cu$_2$ & & \\ \colrule
$J_D$ & $J_\|$ & $J_\perp$ & $J_b$ & $\mathcal{J}_{\alpha}$ & $\mathcal{J}_{\beta}$ 
\\ \colrule
26.5(6) & 3.39(13) & 0.39(3) & $-0.19(2)$ & 0.34(2) & 0.35(2) \\
\end{tabular}
\end{ruledtabular}
\end{table}

The interactions of Table \ref{tab_Jparam} define a Cu$_2$ magnetic lattice 
composed of chains aligned in the ${\bf a} \! - \! {\bf c}$ direction, 
whose energy scale, $J_\|$, exceeds by a factor of 10 all the interchain 
couplings. From Fig.~\ref{fig:structure}(a), $J_\|$ connects Cu$_2$ spins 
through the SeO$_3$ tetrahedra, a superexchange path not so far considered 
\cite{SeCuO3_2012_bulk,SeCuO3_diffraction,secuo3_lee_dec,SeCuO3_decoupled}. 
This coupled-chain character provides an immediate indication for the origin of 
the continuum scattering in Figs.~\ref{lowE_cuts} and \ref{swfitandintensities} 
as the break-up of $\Delta S = 1$ magnons into spinons at energies beyond their 
confinement scale. The four additional Cu$_2$ interactions ensure both strong 
interchain frustration in all three directions and the weak magnetic order at 
$T < T_N$. The small magnon gap can be reproduced with a tiny Ising anisotropy 
of $J_\|$, $\delta J_\| = 0.018$ meV, which has negligible influence on 
the dynamics away from the zone center.
 
To go beyond the independent-subsystem interpretation of the excitation 
spectrum, we restore the coupling between Cu$_1$ and Cu$_2$ atoms in the 
second term of Eq.~(\ref{eq_hamiltonian}), without which the triplon measured 
in Fig.~\ref{highE} would be non-dispersive. All the interactions between the 
Cu$_2$ atoms given in Table \ref{tab_Jparam} are required to fit the dispersion 
data of Fig.~\ref{swfitandintensities} in multiple reciprocal-space directions.
However, they include not only the near-neighbor couplings $J_\|$, $J_\perp$, 
and $J_b$, but also the couplings $\mathcal{J}_{\alpha}$ and $\mathcal{J}_{\beta}$ 
whose long superexchange paths proceed directly across the Cu$_1$--Cu$_1$ dimer 
(Fig.~\ref{fig:structure}); as a result, in a self-consistent model, these 
should be effective couplings arising as a consequence of $J_D$ and of two 
Cu$_1$-Cu$_2$ coupling parameters, $J_{12}^\gamma$ [Fig.~\ref{fig:structure}(e)].

To estimate $J_{12}^\gamma$, we perform a perturbative analysis of the two 
four-site units shown in Fig.~\ref{fig:structure}(e), as detailed in Sec.~S2 
of the SM \cite{sm}. The ground state in the limit $J_{D} \gg J_{12}^\gamma$, 
$\ket{\Phi_0} = \ket{s_1} \otimes \ket{s_2}$, is the product of two singlets 
on each pair of Cu$_1$ and Cu$_2$ sites. In the three lowest excited states, 
the Cu$_1$ dimer remains in a singlet while the Cu$_2$ spins form a triplet, 
$\ket{t_2^l}$, with $l = +, 0, -$. The energy difference gives the effective 
coupling between the two Cu$_2$ atoms,
\begin{equation}\label{eqt_tetramer_jeff}
\mathcal{J}_{\gamma} = \frac{J_{12}^\gamma}{2} + \frac{1}{4} \frac{[ 3 
(J_{12}^\gamma)^2 - 2 J_{D} J_{12}^\gamma]}{\sqrt{J_{D}^2 + (J_{12}^\gamma)^2}} 
\xrightarrow {J_{D} \gg J_{12}^\gamma} \frac{3}{4} \frac{(J_{12}^\gamma)^2}{J_{D}}.
\end{equation}
From the fitted values of the effective couplings $\mathcal{J}_\alpha$ and 
$\mathcal{J}_\beta$ (Table \ref{tab_Jparam}), we deduce the microscopic 
coupling parameters to be $J_{12}^\alpha = 3.47$ meV and $J_{12}^\beta = 3.52$ 
meV. Our LSW fits verify that any effective coupling between atoms connected 
by a path involving both $J_{12}^\alpha$ and $J_{12}^\beta$ must be vanishingly 
small. 

Although these values are large compared to the couplings within the Cu$_2$ 
subsystem in Table \ref{tab_Jparam}, their real effect on the spin dynamics is 
suppressed strongly by $J_D$, as the structure of Eq.~(\ref{eqt_tetramer_jeff}) 
makes clear. Values of 3-4 meV are consistent with the width, 
$\Gamma_\mathcal{L}$, of the triplon at high temperatures [Fig.~\ref{highE}(d)], 
which indicates its coupling to incoherent excitations. The perturbative 
treatment of Eq.~(\ref{eqt_tetramer_jeff}) provides upper bounds for the 
$J_{12}^\gamma$ values and thus is the opposite limit to the LSW approach, which 
cannot describe the full system of Cu$_1$ and Cu$_2$ atoms. The two approaches 
indicate the range of possible renormalization effects due to quantum 
fluctuations, which is widest at intermediate energies (corresponding 
to the spinon continua). The extent of renormalization to LSW theory in 
SeCuO$_3$ can be gauged from the magnetic order on the Cu$_2$ sublattice, 
which despite its 3D nature is $\mu_2 < 0.8 \, \mu_B$ \cite{SeCuO3_diffraction}.
The interactions $J_{12}^\gamma$ induce order on the Cu$_1$ sublattice, although 
$\mu_1 \approx 0.35 \, \mu_B$ is very weak even at the lowest temperatures, and 
hence a full description lies well beyond the LSW approximation. The relative 
canting of the $\mu_1$ and $\mu_2$ moment directions \cite{SeCuO3_2012_bulk} 
and the intensity transfer to the weak magnon branch 
(Fig.~\ref{swfitandintensities}) gauge the physical effects of 
terms omitted in the minimal model of Eq.~(\ref{eq_hamiltonian}).

We return now to the most unexpected feature of our INS data, the 
strong continuum scattering observed directly above the magnon peaks in 
Figs.~\ref{lowE_cuts} and \ref{swfitandintensities}. Interpreting this as 
deconfined spinons requires the definitive exclusion of alternative origins. 
Continuum scattering above a one-magnon band arises naturally due to 
multi-magnon processes and has long been known in both 3D \cite{Cowley1969} 
and 2D antiferromagnets \cite{Huberman2mag}. In this situation, the 
ratio of the integrated intensities, $I_{\rm p}$ in the one-magnon sector and 
$I_{\rm c}$ in the putative multi-magnon sector (Fig.~\ref{swfitandintensities}),
cannot exceed a well-defined limit. In Table \ref{tab_intensities} we average 
both quantities along four $\textbf{q}$ directions, and note further that our 
measured energy range may not capture the upper edge of the continuum 
(Fig.~\ref{lowE_cuts}), whence the ratio $\kappa_{\rm min} = I_{\text{c}}/ 
I_{\text{p}}$ constitutes a lower bound. 

\begin{table}[b]
\caption{\label{tab_intensities}
Integrated peak intensity, $I_{\rm {p}}$, and continuum intensity, $I_{\rm {c}}$, 
averaged along four high-symmetry directions. The lower row presents the LSW 
theoretical result for the spin reduction, $\Delta S_2 = 0.13$, measured 
\cite{SeCuO3_diffraction} on the Cu$_2$ sublattice.}
\begin{ruledtabular}
\begin{tabular}{cccc}
Direction   & $I_{\rm {p}}$ & $I_{\rm {c}}$ & $\kappa_{\text{min}}$ \\ 
\colrule
{[}$h,2,h${]}                       & 90(4)  & 61 & 0.69(4) \\
{[}$0,k,0${]}                       & 133(5) & 99 & 0.85(4) \\
{[}$h,3,h${]}                       & 98(4)  & 125 & 1.28(4) \\
{[}$\frac{1}{2},k,\frac{1}{2}${]} & 85(4)  & 56 & 0.65(4) \\\colrule
$\Delta S_{2} = 0.13$ & 0.47 & 0.15 & 0.32 \\
\end{tabular}
\end{ruledtabular}
\end{table}

We find that $I_{\text{c}}$ is of the same order as $I_{\text{p}}$, making 
their ratio far greater than those found in multimagnon scattering studies 
\cite{Cowley1969,Huberman2mag}. The LSW prediction for this ratio can be 
deduced from the spin reduction (quantum fluctuation renormalization) 
measured \cite{SeCuO3_diffraction} on the Cu$_2$ sublattice, and as Table 
\ref{tab_intensities} makes clear the data exclude any straightforward 
multimagnon origin. We comment that highly unconventional multimagnon 
processes, such as those involving very strong magnon-magnon interactions, 
are hard to rule out, but that there is little evidence for these (such as 
decay of our observed magnons) in SeCuO$_3$. Further, the gap in the one-magnon 
spectrum implies a gap between the one- and multi-magnon contributions, as 
encountered in Ref.~\cite{Huberman2mag}, whereas Figs.~\ref{lowE_cuts} and 
\ref{swfitandintensities} exclude such a gap with even a fraction 
the size of the measured $\Delta$. 

Thus we conclude that the most plausible origin for the observed continuum 
scattering is spinons. More specifically, the strong quantum corrections of 
the chain-like $S = 1/2$ Cu$_2$ system do permit a deconfinement of spinons 
at energies above the one-magnon band. However, the frustrating interchain 
interactions, which allow this spinonic character in a system with magnetic 
order, mean that the resulting continua (Fig.~\ref{swfitandintensities}) 
are far from the familiar single-chain form, which was found in 
Ref.~\cite{Zhang2020} for magnons and spinons coexisting in the quasi-1D 
limit. The problem of partially confined spinons has recently received 
considerable attention in some of the paradigm Heisenberg models within 
frustrated quantum magnetism \cite{Mezio2011,Shao2017,Yu2018,Ghioldi2018,
Ferrari2019}, and SeCuO$_3$ presents a materials example of this complex 
situation. While a detailed analysis lies beyond the scope of the present 
study, the locations of continuum scattering in Fig.~\ref{swfitandintensities} 
and the intensity ratios for magnon and spinon contributions will provide 
essential input for a complete theoretical description. 

To conclude, we have investigated a member of the class of coupled-cluster, 
multi-subsystem quantum magnetic materials in which magnon, triplon, and 
spinon excitations are present simultaneously. In SeCuO$_3$, the clusters 
are strong Cu$_1$ dimers and the second sublattice, Cu$_2$, is a network of 
spin chains on which weak magnetic order appears below $T_N = 8$ K. By 
neutron spectroscopy we have determined not only the intra-sublattice 
interactions but also the Cu$_1$-Cu$_2$ interactions that make the Cu$_1$ 
triplon mode weakly dispersive, induce small Cu$_1$ moments, and create 
frustrating interactions in the Cu$_2$ sublattice, which contribute to the 
emergence of spinon continua above the magnons. From our results, SeCuO$_3$ 
encapsulates the challenge of describing the coherent quantum correlations, 
or quantum entanglement, that in many systems lie beneath the (unentangled) 
veneer of magnetic order, and mandates an integrated theoretical treatment 
of how these correlations lead to all three coexisting excitation types.

\smallskip
\noindent
\begin{acknowledgments}
{\it Acknowledgments.} We thank X. Rocquefelte for helpful discussions. We 
are grateful to the Swiss National Science Foundation (SNSF) for financial 
support under Grant No.~188648 and the European Research Council (ERC) for 
the support of the Synergy Grant HERO. 
Research at TU Dortmund was supported by the Deutsche Forschungsgemeinschaft (DFG, German
Research Foundation) through Grants No. UH90/13-1 and UH90/14-1.
We thank the Institut Laue-Langevin 
and the Laboratoire L\'eon Brillouin for access to their neutron scattering 
facilities. These neutron scattering experiments were performed within the 
framework of the EUROfusion Consortium, which was funded by the Euratom 
Research and Training Programme 2014-2018 under Grant Agreement No.~633053.
\end{acknowledgments}

\bibliography{secuo3bib}

\begin{thebibliography}{45}%
\makeatletter
\providecommand \@ifxundefined [1]{%
 \@ifx{#1\undefined}
}%
\providecommand \@ifnum [1]{%
 \ifnum #1\expandafter \@firstoftwo
 \else \expandafter \@secondoftwo
 \fi
}%
\providecommand \@ifx [1]{%
 \ifx #1\expandafter \@firstoftwo
 \else \expandafter \@secondoftwo
 \fi
}%
\providecommand \natexlab [1]{#1}%
\providecommand \enquote  [1]{``#1''}%
\providecommand \bibnamefont  [1]{#1}%
\providecommand \bibfnamefont [1]{#1}%
\providecommand \citenamefont [1]{#1}%
\providecommand \href@noop [0]{\@secondoftwo}%
\providecommand \href [0]{\begingroup \@sanitize@url \@href}%
\providecommand \@href[1]{\@@startlink{#1}\@@href}%
\providecommand \@@href[1]{\endgroup#1\@@endlink}%
\providecommand \@sanitize@url [0]{\catcode `\\12\catcode `\$12\catcode
  `\&12\catcode `\#12\catcode `\^12\catcode `\_12\catcode `\%12\relax}%
\providecommand \@@startlink[1]{}%
\providecommand \@@endlink[0]{}%
\providecommand \url  [0]{\begingroup\@sanitize@url \@url }%
\providecommand \@url [1]{\endgroup\@href {#1}{\urlprefix }}%
\providecommand \urlprefix  [0]{URL }%
\providecommand \Eprint [0]{\href }%
\providecommand \doibase [0]{http://dx.doi.org/}%
\providecommand \selectlanguage [0]{\@gobble}%
\providecommand \bibinfo  [0]{\@secondoftwo}%
\providecommand \bibfield  [0]{\@secondoftwo}%
\providecommand \translation [1]{[#1]}%
\providecommand \BibitemOpen [0]{}%
\providecommand \bibitemStop [0]{}%
\providecommand \bibitemNoStop [0]{.\EOS\space}%
\providecommand \EOS [0]{\spacefactor3000\relax}%
\providecommand \BibitemShut  [1]{\csname bibitem#1\endcsname}%
\let\auto@bib@innerbib\@empty
\bibitem [{\citenamefont {Cavadini}\ \emph {et~al.}(2001)\citenamefont
  {Cavadini}, \citenamefont {Heigold}, \citenamefont {Henggeler}, \citenamefont
  {Furrer}, \citenamefont {G\"udel}, \citenamefont {Kr\"amer},\ and\
  \citenamefont {Mutka}}]{Cavadini2001}%
  \BibitemOpen
  \bibfield  {author} {\bibinfo {author} {\bibfnamefont {N.}~\bibnamefont
  {Cavadini}}, \bibinfo {author} {\bibfnamefont {G.}~\bibnamefont {Heigold}},
  \bibinfo {author} {\bibfnamefont {W.}~\bibnamefont {Henggeler}}, \bibinfo
  {author} {\bibfnamefont {A.}~\bibnamefont {Furrer}}, \bibinfo {author}
  {\bibfnamefont {H.-U.}\ \bibnamefont {G\"udel}}, \bibinfo {author}
  {\bibfnamefont {K.}~\bibnamefont {Kr\"amer}}, \ and\ \bibinfo {author}
  {\bibfnamefont {H.}~\bibnamefont {Mutka}},\ }\bibfield  {title} {\enquote
  {\bibinfo {title} {{Magnetic excitations in the quantum spin system
  ${\mathrm{TlCuCl}}_{3}$}},}\ }\href {\doibase 10.1103/PhysRevB.63.172414}
  {\bibfield  {journal} {\bibinfo  {journal} {Phys. Rev. B}\ }\textbf {\bibinfo
  {volume} {63}},\ \bibinfo {pages} {172414} (\bibinfo {year}
  {2001})}\BibitemShut {NoStop}%
\bibitem [{\citenamefont {Thielemann}\ \emph {et~al.}(2009)\citenamefont
  {Thielemann}, \citenamefont {R\"uegg}, \citenamefont {R\o{}nnow},
  \citenamefont {L\"auchli}, \citenamefont {Caux}, \citenamefont {Normand},
  \citenamefont {Biner}, \citenamefont {Kr\"amer}, \citenamefont {G\"udel},
  \citenamefont {Stahn}, \citenamefont {Habicht}, \citenamefont {Kiefer},
  \citenamefont {Boehm}, \citenamefont {McMorrow},\ and\ \citenamefont
  {Mesot}}]{Thielemann2009}%
  \BibitemOpen
  \bibfield  {author} {\bibinfo {author} {\bibfnamefont {B.}~\bibnamefont
  {Thielemann}}, \bibinfo {author} {\bibfnamefont {C.}~\bibnamefont {R\"uegg}},
  \bibinfo {author} {\bibfnamefont {H.~M.}\ \bibnamefont {R\o{}nnow}}, \bibinfo
  {author} {\bibfnamefont {A.~M.}\ \bibnamefont {L\"auchli}}, \bibinfo {author}
  {\bibfnamefont {J.-S.}\ \bibnamefont {Caux}}, \bibinfo {author}
  {\bibfnamefont {B.}~\bibnamefont {Normand}}, \bibinfo {author} {\bibfnamefont
  {D.}~\bibnamefont {Biner}}, \bibinfo {author} {\bibfnamefont {K.~W.}\
  \bibnamefont {Kr\"amer}}, \bibinfo {author} {\bibfnamefont {H.-U.}\
  \bibnamefont {G\"udel}}, \bibinfo {author} {\bibfnamefont {J.}~\bibnamefont
  {Stahn}}, \bibinfo {author} {\bibfnamefont {K.}~\bibnamefont {Habicht}},
  \bibinfo {author} {\bibfnamefont {K.}~\bibnamefont {Kiefer}}, \bibinfo
  {author} {\bibfnamefont {M.}~\bibnamefont {Boehm}}, \bibinfo {author}
  {\bibfnamefont {D.~F.}\ \bibnamefont {McMorrow}}, \ and\ \bibinfo {author}
  {\bibfnamefont {J.}~\bibnamefont {Mesot}},\ }\bibfield  {title} {\enquote
  {\bibinfo {title} {{Direct Observation of Magnon Fractionalization in the
  Quantum Spin Ladder}},}\ }\href {\doibase 10.1103/PhysRevLett.102.107204}
  {\bibfield  {journal} {\bibinfo  {journal} {Phys. Rev. Lett.}\ }\textbf
  {\bibinfo {volume} {102}},\ \bibinfo {pages} {107204} (\bibinfo {year}
  {2009})}\BibitemShut {NoStop}%
\bibitem [{\citenamefont {Schmidiger}\ \emph {et~al.}(2013)\citenamefont
  {Schmidiger}, \citenamefont {M\"uhlbauer}, \citenamefont {Zheludev},
  \citenamefont {Bouillot}, \citenamefont {Giamarchi}, \citenamefont {Kollath},
  \citenamefont {Ehlers},\ and\ \citenamefont {Tsvelik}}]{Schmidiger2013}%
  \BibitemOpen
  \bibfield  {author} {\bibinfo {author} {\bibfnamefont {D.}~\bibnamefont
  {Schmidiger}}, \bibinfo {author} {\bibfnamefont {S.}~\bibnamefont
  {M\"uhlbauer}}, \bibinfo {author} {\bibfnamefont {A.}~\bibnamefont
  {Zheludev}}, \bibinfo {author} {\bibfnamefont {P.}~\bibnamefont {Bouillot}},
  \bibinfo {author} {\bibfnamefont {T.}~\bibnamefont {Giamarchi}}, \bibinfo
  {author} {\bibfnamefont {C.}~\bibnamefont {Kollath}}, \bibinfo {author}
  {\bibfnamefont {G.}~\bibnamefont {Ehlers}}, \ and\ \bibinfo {author}
  {\bibfnamefont {A.~M.}\ \bibnamefont {Tsvelik}},\ }\bibfield  {title}
  {\enquote {\bibinfo {title} {Symmetric and asymmetric excitations of a
  strong-leg quantum spin ladder},}\ }\href {\doibase
  10.1103/PhysRevB.88.094411} {\bibfield  {journal} {\bibinfo  {journal} {Phys.
  Rev. B}\ }\textbf {\bibinfo {volume} {88}},\ \bibinfo {pages} {094411}
  (\bibinfo {year} {2013})}\BibitemShut {NoStop}%
\bibitem [{\citenamefont {Gaulin}\ \emph {et~al.}(2004)\citenamefont {Gaulin},
  \citenamefont {Lee}, \citenamefont {Haravifard}, \citenamefont {Castellan},
  \citenamefont {Berlinsky}, \citenamefont {Dabkowska}, \citenamefont {Qiu},\
  and\ \citenamefont {Copley}}]{Gaulin2004}%
  \BibitemOpen
  \bibfield  {author} {\bibinfo {author} {\bibfnamefont {B.~D.}\ \bibnamefont
  {Gaulin}}, \bibinfo {author} {\bibfnamefont {S.~H.}\ \bibnamefont {Lee}},
  \bibinfo {author} {\bibfnamefont {S.}~\bibnamefont {Haravifard}}, \bibinfo
  {author} {\bibfnamefont {J.~P.}\ \bibnamefont {Castellan}}, \bibinfo {author}
  {\bibfnamefont {A.~J.}\ \bibnamefont {Berlinsky}}, \bibinfo {author}
  {\bibfnamefont {H.~A.}\ \bibnamefont {Dabkowska}}, \bibinfo {author}
  {\bibfnamefont {Y.}~\bibnamefont {Qiu}}, \ and\ \bibinfo {author}
  {\bibfnamefont {J.~R.~D.}\ \bibnamefont {Copley}},\ }\bibfield  {title}
  {\enquote {\bibinfo {title} {{High-Resolution Study of Spin Excitations in
  the Singlet Ground State of
  ${\mathrm{S}\mathrm{r}\mathrm{C}\mathrm{u}}_{2}({\mathrm{B}\mathrm{O}}_{3}{)}_{2}$}},}\
  }\href {\doibase 10.1103/PhysRevLett.93.267202} {\bibfield  {journal}
  {\bibinfo  {journal} {Phys. Rev. Lett.}\ }\textbf {\bibinfo {volume} {93}},\
  \bibinfo {pages} {267202} (\bibinfo {year} {2004})}\BibitemShut {NoStop}%
\bibitem [{\citenamefont {Torrance}\ and\ \citenamefont
  {Tinkham}(1969)}]{Torrance1969}%
  \BibitemOpen
  \bibfield  {author} {\bibinfo {author} {\bibfnamefont {J.~B.}\ \bibnamefont
  {Torrance}}\ and\ \bibinfo {author} {\bibfnamefont {M.}~\bibnamefont
  {Tinkham}},\ }\bibfield  {title} {\enquote {\bibinfo {title} {{Excitation of
  Multiple-Magnon Bound States in
  Co${\mathrm{Cl}}_{2}$\ifmmode\cdot\else\textperiodcentered\fi{}2${\mathrm{H}}_{2}$O}},}\
  }\href {\doibase 10.1103/PhysRev.187.595} {\bibfield  {journal} {\bibinfo
  {journal} {Phys. Rev.}\ }\textbf {\bibinfo {volume} {187}},\ \bibinfo {pages}
  {595--606} (\bibinfo {year} {1969})}\BibitemShut {NoStop}%
\bibitem [{\citenamefont {Uhrig}\ and\ \citenamefont
  {Schulz}(1996)}]{Uhrig1996}%
  \BibitemOpen
  \bibfield  {author} {\bibinfo {author} {\bibfnamefont {G.~S.}\ \bibnamefont
  {Uhrig}}\ and\ \bibinfo {author} {\bibfnamefont {H.~J.}\ \bibnamefont
  {Schulz}},\ }\bibfield  {title} {\enquote {\bibinfo {title} {Magnetic
  excitation spectrum of dimerized antiferromagnetic chains},}\ }\href
  {\doibase 10.1103/PhysRevB.54.R9624} {\bibfield  {journal} {\bibinfo
  {journal} {Phys. Rev. B}\ }\textbf {\bibinfo {volume} {54}},\ \bibinfo
  {pages} {R9624--R9627} (\bibinfo {year} {1996})}\BibitemShut {NoStop}%
\bibitem [{\citenamefont {Ward}\ \emph {et~al.}(2017)\citenamefont {Ward},
  \citenamefont {Mena}, \citenamefont {Bouillot}, \citenamefont {Kollath},
  \citenamefont {Giamarchi}, \citenamefont {Schmidt}, \citenamefont {Normand},
  \citenamefont {Kr\"amer}, \citenamefont {Biner}, \citenamefont {Bewley},
  \citenamefont {Guidi}, \citenamefont {Boehm}, \citenamefont {McMorrow},\ and\
  \citenamefont {R\"uegg}}]{Ward2017}%
  \BibitemOpen
  \bibfield  {author} {\bibinfo {author} {\bibfnamefont {S.}~\bibnamefont
  {Ward}}, \bibinfo {author} {\bibfnamefont {M.}~\bibnamefont {Mena}}, \bibinfo
  {author} {\bibfnamefont {P.}~\bibnamefont {Bouillot}}, \bibinfo {author}
  {\bibfnamefont {C.}~\bibnamefont {Kollath}}, \bibinfo {author} {\bibfnamefont
  {T.}~\bibnamefont {Giamarchi}}, \bibinfo {author} {\bibfnamefont {K.~P.}\
  \bibnamefont {Schmidt}}, \bibinfo {author} {\bibfnamefont {B.}~\bibnamefont
  {Normand}}, \bibinfo {author} {\bibfnamefont {K.~W.}\ \bibnamefont
  {Kr\"amer}}, \bibinfo {author} {\bibfnamefont {D.}~\bibnamefont {Biner}},
  \bibinfo {author} {\bibfnamefont {R.}~\bibnamefont {Bewley}}, \bibinfo
  {author} {\bibfnamefont {T.}~\bibnamefont {Guidi}}, \bibinfo {author}
  {\bibfnamefont {M.}~\bibnamefont {Boehm}}, \bibinfo {author} {\bibfnamefont
  {D.~F.}\ \bibnamefont {McMorrow}}, \ and\ \bibinfo {author} {\bibfnamefont
  {C.}~\bibnamefont {R\"uegg}},\ }\bibfield  {title} {\enquote {\bibinfo
  {title} {{Bound States and Field-Polarized Haldane Modes in a Quantum Spin
  Ladder}},}\ }\href {\doibase 10.1103/PhysRevLett.118.177202} {\bibfield
  {journal} {\bibinfo  {journal} {Phys. Rev. Lett.}\ }\textbf {\bibinfo
  {volume} {118}},\ \bibinfo {pages} {177202} (\bibinfo {year}
  {2017})}\BibitemShut {NoStop}%
\bibitem [{\citenamefont {McClarty}\ \emph {et~al.}(2017)\citenamefont
  {McClarty}, \citenamefont {Kr{\"u}ger}, \citenamefont {Guidi}, \citenamefont
  {Parker}, \citenamefont {Refson}, \citenamefont {Parker}, \citenamefont
  {Prabhakaran},\ and\ \citenamefont {Coldea}}]{McClarty2017}%
  \BibitemOpen
  \bibfield  {author} {\bibinfo {author} {\bibfnamefont {P.~A.}\ \bibnamefont
  {McClarty}}, \bibinfo {author} {\bibfnamefont {F.}~\bibnamefont
  {Kr{\"u}ger}}, \bibinfo {author} {\bibfnamefont {T.}~\bibnamefont {Guidi}},
  \bibinfo {author} {\bibfnamefont {S.~F.}\ \bibnamefont {Parker}}, \bibinfo
  {author} {\bibfnamefont {K.}~\bibnamefont {Refson}}, \bibinfo {author}
  {\bibfnamefont {A.~W.}\ \bibnamefont {Parker}}, \bibinfo {author}
  {\bibfnamefont {D.}~\bibnamefont {Prabhakaran}}, \ and\ \bibinfo {author}
  {\bibfnamefont {R.}~\bibnamefont {Coldea}},\ }\bibfield  {title} {\enquote
  {\bibinfo {title} {{Topological triplon modes and bound states in a
  Shastry--Sutherland magnet}},}\ }\href {\doibase 10.1038/nphys4117}
  {\bibfield  {journal} {\bibinfo  {journal} {Nat. Phys.}\ }\textbf {\bibinfo
  {volume} {13}},\ \bibinfo {pages} {736--741} (\bibinfo {year}
  {2017})}\BibitemShut {NoStop}%
\bibitem [{\citenamefont {Hammar}\ \emph {et~al.}(1999)\citenamefont {Hammar},
  \citenamefont {Stone}, \citenamefont {Reich}, \citenamefont {Broholm},
  \citenamefont {Gibson}, \citenamefont {Turnbull}, \citenamefont {Landee},\
  and\ \citenamefont {Oshikawa}}]{Hammar1999}%
  \BibitemOpen
  \bibfield  {author} {\bibinfo {author} {\bibfnamefont {P.~R.}\ \bibnamefont
  {Hammar}}, \bibinfo {author} {\bibfnamefont {M.~B.}\ \bibnamefont {Stone}},
  \bibinfo {author} {\bibfnamefont {D.~H.}\ \bibnamefont {Reich}}, \bibinfo
  {author} {\bibfnamefont {C.}~\bibnamefont {Broholm}}, \bibinfo {author}
  {\bibfnamefont {P.~J.}\ \bibnamefont {Gibson}}, \bibinfo {author}
  {\bibfnamefont {M.~M.}\ \bibnamefont {Turnbull}}, \bibinfo {author}
  {\bibfnamefont {C.~P.}\ \bibnamefont {Landee}}, \ and\ \bibinfo {author}
  {\bibfnamefont {M.}~\bibnamefont {Oshikawa}},\ }\bibfield  {title} {\enquote
  {\bibinfo {title} {{Characterization of a quasi-one-dimensional spin-1/2
  magnet which is gapless and paramagnetic for
  $g{\ensuremath{\mu}}_{B}H\ensuremath{\lesssim}J$ and
  ${k}_{B}T\ensuremath{\ll}J$}},}\ }\href {\doibase 10.1103/PhysRevB.59.1008}
  {\bibfield  {journal} {\bibinfo  {journal} {Phys. Rev. B}\ }\textbf {\bibinfo
  {volume} {59}},\ \bibinfo {pages} {1008--1015} (\bibinfo {year}
  {1999})}\BibitemShut {NoStop}%
\bibitem [{\citenamefont {Mourigal}\ \emph {et~al.}(2013)\citenamefont
  {Mourigal}, \citenamefont {Enderle}, \citenamefont {Kl{\"o}pperpieper},
  \citenamefont {Caux}, \citenamefont {Stunault},\ and\ \citenamefont
  {R{\o}nnow}}]{Mourigal2013}%
  \BibitemOpen
  \bibfield  {author} {\bibinfo {author} {\bibfnamefont {M.}~\bibnamefont
  {Mourigal}}, \bibinfo {author} {\bibfnamefont {M.}~\bibnamefont {Enderle}},
  \bibinfo {author} {\bibfnamefont {A.}~\bibnamefont {Kl{\"o}pperpieper}},
  \bibinfo {author} {\bibfnamefont {J.-S.}\ \bibnamefont {Caux}}, \bibinfo
  {author} {\bibfnamefont {A.}~\bibnamefont {Stunault}}, \ and\ \bibinfo
  {author} {\bibfnamefont {H.~M.}\ \bibnamefont {R{\o}nnow}},\ }\bibfield
  {title} {\enquote {\bibinfo {title} {Fractional spinon excitations in the
  quantum {H}eisenberg antiferromagnetic chain},}\ }\href {\doibase
  10.1038/nphys2652} {\bibfield  {journal} {\bibinfo  {journal} {Nat. Phys.}\
  }\textbf {\bibinfo {volume} {9}},\ \bibinfo {pages} {435--441} (\bibinfo
  {year} {2013})}\BibitemShut {NoStop}%
\bibitem [{\citenamefont {Nagler}\ \emph {et~al.}(1983)\citenamefont {Nagler},
  \citenamefont {Buyers}, \citenamefont {Armstrong},\ and\ \citenamefont
  {Briat}}]{Nagler1983}%
  \BibitemOpen
  \bibfield  {author} {\bibinfo {author} {\bibfnamefont {S.~E.}\ \bibnamefont
  {Nagler}}, \bibinfo {author} {\bibfnamefont {W.~J.~L.}\ \bibnamefont
  {Buyers}}, \bibinfo {author} {\bibfnamefont {R.~L.}\ \bibnamefont
  {Armstrong}}, \ and\ \bibinfo {author} {\bibfnamefont {B.}~\bibnamefont
  {Briat}},\ }\bibfield  {title} {\enquote {\bibinfo {title} {{Solitons in the
  one-dimensional antiferromagnet CsCo${\mathrm{Br}}_{3}$}},}\ }\href {\doibase
  10.1103/PhysRevB.28.3873} {\bibfield  {journal} {\bibinfo  {journal} {Phys.
  Rev. B}\ }\textbf {\bibinfo {volume} {28}},\ \bibinfo {pages} {3873--3885}
  (\bibinfo {year} {1983})}\BibitemShut {NoStop}%
\bibitem [{\citenamefont {Faure}\ \emph {et~al.}(2018)\citenamefont {Faure},
  \citenamefont {Takayoshi}, \citenamefont {Petit}, \citenamefont {Simonet},
  \citenamefont {Raymond}, \citenamefont {Regnault}, \citenamefont {Boehm},
  \citenamefont {White}, \citenamefont {M{\aa}nsson}, \citenamefont
  {R{\"u}egg}, \citenamefont {Lejay}, \citenamefont {Canals}, \citenamefont
  {Lorenz}, \citenamefont {Furuya}, \citenamefont {Giamarchi},\ and\
  \citenamefont {Grenier}}]{Faure2018}%
  \BibitemOpen
  \bibfield  {author} {\bibinfo {author} {\bibfnamefont {Q.}~\bibnamefont
  {Faure}}, \bibinfo {author} {\bibfnamefont {S.}~\bibnamefont {Takayoshi}},
  \bibinfo {author} {\bibfnamefont {S.}~\bibnamefont {Petit}}, \bibinfo
  {author} {\bibfnamefont {V.}~\bibnamefont {Simonet}}, \bibinfo {author}
  {\bibfnamefont {S.}~\bibnamefont {Raymond}}, \bibinfo {author} {\bibfnamefont
  {L.-P.}\ \bibnamefont {Regnault}}, \bibinfo {author} {\bibfnamefont
  {M.}~\bibnamefont {Boehm}}, \bibinfo {author} {\bibfnamefont {J.~S.}\
  \bibnamefont {White}}, \bibinfo {author} {\bibfnamefont {M.}~\bibnamefont
  {M{\aa}nsson}}, \bibinfo {author} {\bibfnamefont {C.}~\bibnamefont
  {R{\"u}egg}}, \bibinfo {author} {\bibfnamefont {P.}~\bibnamefont {Lejay}},
  \bibinfo {author} {\bibfnamefont {B.}~\bibnamefont {Canals}}, \bibinfo
  {author} {\bibfnamefont {T.}~\bibnamefont {Lorenz}}, \bibinfo {author}
  {\bibfnamefont {S.~C.}\ \bibnamefont {Furuya}}, \bibinfo {author}
  {\bibfnamefont {T.}~\bibnamefont {Giamarchi}}, \ and\ \bibinfo {author}
  {\bibfnamefont {B.}~\bibnamefont {Grenier}},\ }\bibfield  {title} {\enquote
  {\bibinfo {title} {Topological quantum phase transition in the {I}sing-like
  antiferromagnetic spin chain {BaCo$_2$V$_2$O$_8$}},}\ }\href {\doibase
  10.1038/s41567-018-0126-8} {\bibfield  {journal} {\bibinfo  {journal} {Nat.
  Phys.}\ }\textbf {\bibinfo {volume} {14}},\ \bibinfo {pages} {716--722}
  (\bibinfo {year} {2018})}\BibitemShut {NoStop}%
\bibitem [{\citenamefont {Kitaev}(2006)}]{Kitaev2006}%
  \BibitemOpen
  \bibfield  {author} {\bibinfo {author} {\bibfnamefont {A.}~\bibnamefont
  {Kitaev}},\ }\bibfield  {title} {\enquote {\bibinfo {title} {Anyons in an
  exactly solved model and beyond},}\ }\href {\doibase
  https://doi.org/10.1016/j.aop.2005.10.005} {\bibfield  {journal} {\bibinfo
  {journal} {Ann. Phys.}\ }\textbf {\bibinfo {volume} {321}},\ \bibinfo {pages}
  {2--111} (\bibinfo {year} {2006})}\BibitemShut {NoStop}%
\bibitem [{\citenamefont {Chaloupka}\ \emph {et~al.}(2010)\citenamefont
  {Chaloupka}, \citenamefont {Jackeli},\ and\ \citenamefont
  {Khaliullin}}]{Chaloupka2010}%
  \BibitemOpen
  \bibfield  {author} {\bibinfo {author} {\bibfnamefont {J.}~\bibnamefont
  {Chaloupka}}, \bibinfo {author} {\bibfnamefont {G.}~\bibnamefont {Jackeli}},
  \ and\ \bibinfo {author} {\bibfnamefont {G.}~\bibnamefont {Khaliullin}},\
  }\bibfield  {title} {\enquote {\bibinfo {title} {{Kitaev-Heisenberg Model on
  a Honeycomb Lattice: Possible Exotic Phases in Iridium Oxides
  ${A}_{2}{\mathrm{IrO}}_{3}$}},}\ }\href {\doibase
  10.1103/PhysRevLett.105.027204} {\bibfield  {journal} {\bibinfo  {journal}
  {Phys. Rev. Lett.}\ }\textbf {\bibinfo {volume} {105}},\ \bibinfo {pages}
  {027204} (\bibinfo {year} {2010})}\BibitemShut {NoStop}%
\bibitem [{\citenamefont {Wang}\ \emph {et~al.}(2019)\citenamefont {Wang},
  \citenamefont {Normand},\ and\ \citenamefont {Liu}}]{Wang2019}%
  \BibitemOpen
  \bibfield  {author} {\bibinfo {author} {\bibfnamefont {J.}~\bibnamefont
  {Wang}}, \bibinfo {author} {\bibfnamefont {B.}~\bibnamefont {Normand}}, \
  and\ \bibinfo {author} {\bibfnamefont {Z.-X.}\ \bibnamefont {Liu}},\
  }\bibfield  {title} {\enquote {\bibinfo {title} {{One Proximate Kitaev Spin
  Liquid in the $K$-$J$-$\Gamma$ Model on the Honeycomb Lattice}},}\ }\href
  {\doibase 10.1103/PhysRevLett.123.197201} {\bibfield  {journal} {\bibinfo
  {journal} {Phys. Rev. Lett.}\ }\textbf {\bibinfo {volume} {123}},\ \bibinfo
  {pages} {197201} (\bibinfo {year} {2019})}\BibitemShut {NoStop}%
\bibitem [{\citenamefont {R{\"u}egg}\ \emph {et~al.}(2003)\citenamefont
  {R{\"u}egg}, \citenamefont {Cavadini}, \citenamefont {Furrer}, \citenamefont
  {G{\"u}del}, \citenamefont {Kr{\"a}mer}, \citenamefont {Mutka}, \citenamefont
  {Wildes}, \citenamefont {Habicht},\ and\ \citenamefont
  {Vorderwisch}}]{Rueegg2003}%
  \BibitemOpen
  \bibfield  {author} {\bibinfo {author} {\bibfnamefont {C.}~\bibnamefont
  {R{\"u}egg}}, \bibinfo {author} {\bibfnamefont {N.}~\bibnamefont {Cavadini}},
  \bibinfo {author} {\bibfnamefont {A.}~\bibnamefont {Furrer}}, \bibinfo
  {author} {\bibfnamefont {H.-U.}\ \bibnamefont {G{\"u}del}}, \bibinfo {author}
  {\bibfnamefont {K.}~\bibnamefont {Kr{\"a}mer}}, \bibinfo {author}
  {\bibfnamefont {H.}~\bibnamefont {Mutka}}, \bibinfo {author} {\bibfnamefont
  {A.}~\bibnamefont {Wildes}}, \bibinfo {author} {\bibfnamefont
  {K.}~\bibnamefont {Habicht}}, \ and\ \bibinfo {author} {\bibfnamefont
  {P.}~\bibnamefont {Vorderwisch}},\ }\bibfield  {title} {\enquote {\bibinfo
  {title} {{Bose--Einstein condensation of the triplet states in the magnetic
  insulator TlCuCl$_3$}},}\ }\href {\doibase 10.1038/nature01617} {\bibfield
  {journal} {\bibinfo  {journal} {Nature}\ }\textbf {\bibinfo {volume} {423}},\
  \bibinfo {pages} {62--65} (\bibinfo {year} {2003})}\BibitemShut {NoStop}%
\bibitem [{\citenamefont {R\"uegg}\ \emph {et~al.}(2008)\citenamefont
  {R\"uegg}, \citenamefont {Normand}, \citenamefont {Matsumoto}, \citenamefont
  {Furrer}, \citenamefont {McMorrow}, \citenamefont {Kr\"amer}, \citenamefont
  {G\"udel}, \citenamefont {Gvasaliya}, \citenamefont {Mutka},\ and\
  \citenamefont {Boehm}}]{Rueegg2008}%
  \BibitemOpen
  \bibfield  {author} {\bibinfo {author} {\bibfnamefont {C.}~\bibnamefont
  {R\"uegg}}, \bibinfo {author} {\bibfnamefont {B.}~\bibnamefont {Normand}},
  \bibinfo {author} {\bibfnamefont {M.}~\bibnamefont {Matsumoto}}, \bibinfo
  {author} {\bibfnamefont {A.}~\bibnamefont {Furrer}}, \bibinfo {author}
  {\bibfnamefont {D.~F.}\ \bibnamefont {McMorrow}}, \bibinfo {author}
  {\bibfnamefont {K.~W.}\ \bibnamefont {Kr\"amer}}, \bibinfo {author}
  {\bibfnamefont {H.~U.}\ \bibnamefont {G\"udel}}, \bibinfo {author}
  {\bibfnamefont {S.~N.}\ \bibnamefont {Gvasaliya}}, \bibinfo {author}
  {\bibfnamefont {H.}~\bibnamefont {Mutka}}, \ and\ \bibinfo {author}
  {\bibfnamefont {M.}~\bibnamefont {Boehm}},\ }\bibfield  {title} {\enquote
  {\bibinfo {title} {{Quantum Magnets under Pressure: Controlling Elementary
  Excitations in ${\mathrm{TlCuCl}}_{3}$}},}\ }\href {\doibase
  10.1103/PhysRevLett.100.205701} {\bibfield  {journal} {\bibinfo  {journal}
  {Phys. Rev. Lett.}\ }\textbf {\bibinfo {volume} {100}},\ \bibinfo {pages}
  {205701} (\bibinfo {year} {2008})}\BibitemShut {NoStop}%
\bibitem [{\citenamefont {Lake}\ \emph {et~al.}(2005)\citenamefont {Lake},
  \citenamefont {Tennant}, \citenamefont {Frost},\ and\ \citenamefont
  {Nagler}}]{Lake2005}%
  \BibitemOpen
  \bibfield  {author} {\bibinfo {author} {\bibfnamefont {B.}~\bibnamefont
  {Lake}}, \bibinfo {author} {\bibfnamefont {D.~A.}\ \bibnamefont {Tennant}},
  \bibinfo {author} {\bibfnamefont {C.~D.}\ \bibnamefont {Frost}}, \ and\
  \bibinfo {author} {\bibfnamefont {S.~E.}\ \bibnamefont {Nagler}},\ }\bibfield
   {title} {\enquote {\bibinfo {title} {Quantum criticality and universal
  scaling of a quantum antiferromagnet},}\ }\href {\doibase 10.1038/nmat1327}
  {\bibfield  {journal} {\bibinfo  {journal} {Nat. Mater.}\ }\textbf {\bibinfo
  {volume} {4}},\ \bibinfo {pages} {329--334} (\bibinfo {year}
  {2005})}\BibitemShut {NoStop}%
\bibitem [{\citenamefont {Dalla~Piazza}\ \emph {et~al.}(2015)\citenamefont
  {Dalla~Piazza}, \citenamefont {Mourigal}, \citenamefont {Christensen},
  \citenamefont {Nilsen}, \citenamefont {Tregenna-Piggott}, \citenamefont
  {Perring}, \citenamefont {Enderle}, \citenamefont {McMorrow}, \citenamefont
  {Ivanov},\ and\ \citenamefont {R{\o}nnow}}]{DallaPiazza2015}%
  \BibitemOpen
  \bibfield  {author} {\bibinfo {author} {\bibfnamefont {B.}~\bibnamefont
  {Dalla~Piazza}}, \bibinfo {author} {\bibfnamefont {M.}~\bibnamefont
  {Mourigal}}, \bibinfo {author} {\bibfnamefont {N.~B.}\ \bibnamefont
  {Christensen}}, \bibinfo {author} {\bibfnamefont {G.~J.}\ \bibnamefont
  {Nilsen}}, \bibinfo {author} {\bibfnamefont {P.}~\bibnamefont
  {Tregenna-Piggott}}, \bibinfo {author} {\bibfnamefont {T.~G.}\ \bibnamefont
  {Perring}}, \bibinfo {author} {\bibfnamefont {M.}~\bibnamefont {Enderle}},
  \bibinfo {author} {\bibfnamefont {D.~F.}\ \bibnamefont {McMorrow}}, \bibinfo
  {author} {\bibfnamefont {D.~A.}\ \bibnamefont {Ivanov}}, \ and\ \bibinfo
  {author} {\bibfnamefont {H.~M.}\ \bibnamefont {R{\o}nnow}},\ }\bibfield
  {title} {\enquote {\bibinfo {title} {Fractional excitations in the
  square-lattice quantum antiferromagnet},}\ }\href {\doibase
  10.1038/nphys3172} {\bibfield  {journal} {\bibinfo  {journal} {Nat. Phys.}\
  }\textbf {\bibinfo {volume} {11}},\ \bibinfo {pages} {62--68} (\bibinfo
  {year} {2015})}\BibitemShut {NoStop}%
\bibitem [{\citenamefont {Skoulatos}\ \emph {et~al.}(2017)\citenamefont
  {Skoulatos}, \citenamefont {M\aa{}nsson}, \citenamefont {Fiolka},
  \citenamefont {Kr\"amer}, \citenamefont {Schefer}, \citenamefont {White},\
  and\ \citenamefont {R\"uegg}}]{Skoulatos2017}%
  \BibitemOpen
  \bibfield  {author} {\bibinfo {author} {\bibfnamefont {M.}~\bibnamefont
  {Skoulatos}}, \bibinfo {author} {\bibfnamefont {M.}~\bibnamefont
  {M\aa{}nsson}}, \bibinfo {author} {\bibfnamefont {C.}~\bibnamefont {Fiolka}},
  \bibinfo {author} {\bibfnamefont {K.~W.}\ \bibnamefont {Kr\"amer}}, \bibinfo
  {author} {\bibfnamefont {J.}~\bibnamefont {Schefer}}, \bibinfo {author}
  {\bibfnamefont {J.~S.}\ \bibnamefont {White}}, \ and\ \bibinfo {author}
  {\bibfnamefont {C.}~\bibnamefont {R\"uegg}},\ }\bibfield  {title} {\enquote
  {\bibinfo {title} {{Dimensional reduction by pressure in the magnetic
  framework material
  ${\mathrm{CuF}}_{2}{({\mathrm{D}}_{2}\mathrm{O})}_{2}$(pyz): From spin-wave
  to spinon excitations}},}\ }\href {\doibase 10.1103/PhysRevB.96.020414}
  {\bibfield  {journal} {\bibinfo  {journal} {Phys. Rev. B}\ }\textbf {\bibinfo
  {volume} {96}},\ \bibinfo {pages} {020414} (\bibinfo {year}
  {2017})}\BibitemShut {NoStop}%
\bibitem [{\citenamefont {Johnsson}\ \emph {et~al.}(2000)\citenamefont
  {Johnsson}, \citenamefont {T{\"o}rnroos}, \citenamefont {Mila},\ and\
  \citenamefont {Millet}}]{Johnsson2000}%
  \BibitemOpen
  \bibfield  {author} {\bibinfo {author} {\bibfnamefont {M.}~\bibnamefont
  {Johnsson}}, \bibinfo {author} {\bibfnamefont {K.~W.}\ \bibnamefont
  {T{\"o}rnroos}}, \bibinfo {author} {\bibfnamefont {F.}~\bibnamefont {Mila}},
  \ and\ \bibinfo {author} {\bibfnamefont {P.}~\bibnamefont {Millet}},\
  }\bibfield  {title} {\enquote {\bibinfo {title} {{Tetrahedral Clusters of
  Copper(II): Crystal Structures and Magnetic Properties of
  Cu$_2$Te$_2$O$_5$X$_2$ (X = Cl, Br)}},}\ }\href {\doibase 10.1021/cm000218k}
  {\bibfield  {journal} {\bibinfo  {journal} {Chem. Mater.}\ }\textbf {\bibinfo
  {volume} {12}},\ \bibinfo {pages} {2853--2857} (\bibinfo {year}
  {2000})}\BibitemShut {NoStop}%
\bibitem [{\citenamefont {Jagli\ifmmode \check{c}\else
  \v{c}\fi{}i\ifmmode~\acute{c}\else \'{c}\fi{}}\ \emph
  {et~al.}(2006)\citenamefont {Jagli\ifmmode \check{c}\else
  \v{c}\fi{}i\ifmmode~\acute{c}\else \'{c}\fi{}}, \citenamefont {El~Shawish},
  \citenamefont {Jeromen}, \citenamefont {Bilu\ifmmode \check{s}\else
  \v{s}\fi{}i\ifmmode~\acute{c}\else \'{c}\fi{}}, \citenamefont {Smontara},
  \citenamefont {Trontelj}, \citenamefont {Bon\ifmmode~\check{c}\else
  \v{c}\fi{}a}, \citenamefont {Dolin\ifmmode~\check{s}\else \v{s}\fi{}ek},\
  and\ \citenamefont {Berger}}]{Jaglicic2006}%
  \BibitemOpen
  \bibfield  {author} {\bibinfo {author} {\bibfnamefont {Z.}~\bibnamefont
  {Jagli\ifmmode \check{c}\else \v{c}\fi{}i\ifmmode~\acute{c}\else
  \'{c}\fi{}}}, \bibinfo {author} {\bibfnamefont {S.}~\bibnamefont
  {El~Shawish}}, \bibinfo {author} {\bibfnamefont {A.}~\bibnamefont {Jeromen}},
  \bibinfo {author} {\bibfnamefont {A.}~\bibnamefont {Bilu\ifmmode
  \check{s}\else \v{s}\fi{}i\ifmmode~\acute{c}\else \'{c}\fi{}}}, \bibinfo
  {author} {\bibfnamefont {A.}~\bibnamefont {Smontara}}, \bibinfo {author}
  {\bibfnamefont {Z.}~\bibnamefont {Trontelj}}, \bibinfo {author}
  {\bibfnamefont {J.}~\bibnamefont {Bon\ifmmode~\check{c}\else \v{c}\fi{}a}},
  \bibinfo {author} {\bibfnamefont {J.}~\bibnamefont
  {Dolin\ifmmode~\check{s}\else \v{s}\fi{}ek}}, \ and\ \bibinfo {author}
  {\bibfnamefont {H.}~\bibnamefont {Berger}},\ }\bibfield  {title} {\enquote
  {\bibinfo {title} {{Magnetic ordering and ergodicity of the spin system in
  the ${\mathrm{Cu}}_{2}{\mathrm{Te}}_{2}{\mathrm{O}}_{5}{X}_{2}$ family of
  quantum magnets}},}\ }\href {\doibase 10.1103/PhysRevB.73.214408} {\bibfield
  {journal} {\bibinfo  {journal} {Phys. Rev. B}\ }\textbf {\bibinfo {volume}
  {73}},\ \bibinfo {pages} {214408} (\bibinfo {year} {2006})}\BibitemShut
  {NoStop}%
\bibitem [{\citenamefont {Kotov}\ \emph {et~al.}(2004)\citenamefont {Kotov},
  \citenamefont {Zhitomirsky}, \citenamefont {Elhajal},\ and\ \citenamefont
  {Mila}}]{Kotov2004}%
  \BibitemOpen
  \bibfield  {author} {\bibinfo {author} {\bibfnamefont {V.~N.}\ \bibnamefont
  {Kotov}}, \bibinfo {author} {\bibfnamefont {M.~E.}\ \bibnamefont
  {Zhitomirsky}}, \bibinfo {author} {\bibfnamefont {M.}~\bibnamefont
  {Elhajal}}, \ and\ \bibinfo {author} {\bibfnamefont {F.}~\bibnamefont
  {Mila}},\ }\bibfield  {title} {\enquote {\bibinfo {title} {Weak
  antiferromagnetism and dimer order in quantum systems of coupled
  tetrahedra},}\ }\href {\doibase 10.1103/PhysRevB.70.214401} {\bibfield
  {journal} {\bibinfo  {journal} {Phys. Rev. B}\ }\textbf {\bibinfo {volume}
  {70}},\ \bibinfo {pages} {214401} (\bibinfo {year} {2004})}\BibitemShut
  {NoStop}%
\bibitem [{\citenamefont {Zaharko}\ \emph {et~al.}(2004)\citenamefont
  {Zaharko}, \citenamefont {Daoud-Aladine}, \citenamefont {Streule},
  \citenamefont {Mesot}, \citenamefont {Brown},\ and\ \citenamefont
  {Berger}}]{Zaharko2004}%
  \BibitemOpen
  \bibfield  {author} {\bibinfo {author} {\bibfnamefont {O.}~\bibnamefont
  {Zaharko}}, \bibinfo {author} {\bibfnamefont {A.}~\bibnamefont
  {Daoud-Aladine}}, \bibinfo {author} {\bibfnamefont {S.}~\bibnamefont
  {Streule}}, \bibinfo {author} {\bibfnamefont {J.}~\bibnamefont {Mesot}},
  \bibinfo {author} {\bibfnamefont {P.-J.}\ \bibnamefont {Brown}}, \ and\
  \bibinfo {author} {\bibfnamefont {H.}~\bibnamefont {Berger}},\ }\bibfield
  {title} {\enquote {\bibinfo {title} {{Incommensurate Magnetic Ordering in
  ${\mathrm{C}\mathrm{u}}_{2}{\mathrm{T}\mathrm{e}}_{2}{\mathrm{O}}_{5}{X}_{2}$
  ($X = \mathrm{C}\mathrm{l}, \, \mathrm{B}\mathrm{r}$) Studied by Neutron
  Diffraction}},}\ }\href {\doibase 10.1103/PhysRevLett.93.217206} {\bibfield
  {journal} {\bibinfo  {journal} {Phys. Rev. Lett.}\ }\textbf {\bibinfo
  {volume} {93}},\ \bibinfo {pages} {217206} (\bibinfo {year}
  {2004})}\BibitemShut {NoStop}%
\bibitem [{\citenamefont {Pr\ifmmode~\check{s}\else \v{s}\fi{}a}\ \emph
  {et~al.}(2009)\citenamefont {Pr\ifmmode~\check{s}\else \v{s}\fi{}a},
  \citenamefont {R\o{}nnow}, \citenamefont {Zaharko}, \citenamefont
  {Christensen}, \citenamefont {Jensen}, \citenamefont {Chang}, \citenamefont
  {Streule}, \citenamefont {Jim\'enez-Ruiz}, \citenamefont {Berger},
  \citenamefont {Prester},\ and\ \citenamefont {Mesot}}]{Prsa2009}%
  \BibitemOpen
  \bibfield  {author} {\bibinfo {author} {\bibfnamefont {K.}~\bibnamefont
  {Pr\ifmmode~\check{s}\else \v{s}\fi{}a}}, \bibinfo {author} {\bibfnamefont
  {H.~M.}\ \bibnamefont {R\o{}nnow}}, \bibinfo {author} {\bibfnamefont
  {O.}~\bibnamefont {Zaharko}}, \bibinfo {author} {\bibfnamefont {N.~B.}\
  \bibnamefont {Christensen}}, \bibinfo {author} {\bibfnamefont
  {J.}~\bibnamefont {Jensen}}, \bibinfo {author} {\bibfnamefont
  {J.}~\bibnamefont {Chang}}, \bibinfo {author} {\bibfnamefont
  {S.}~\bibnamefont {Streule}}, \bibinfo {author} {\bibfnamefont
  {M.}~\bibnamefont {Jim\'enez-Ruiz}}, \bibinfo {author} {\bibfnamefont
  {H.}~\bibnamefont {Berger}}, \bibinfo {author} {\bibfnamefont
  {M.}~\bibnamefont {Prester}}, \ and\ \bibinfo {author} {\bibfnamefont
  {J.}~\bibnamefont {Mesot}},\ }\bibfield  {title} {\enquote {\bibinfo {title}
  {{Anomalous Magnetic Excitations of Cooperative Tetrahedral Spin
  Clusters}},}\ }\href {\doibase 10.1103/PhysRevLett.102.177202} {\bibfield
  {journal} {\bibinfo  {journal} {Phys. Rev. Lett.}\ }\textbf {\bibinfo
  {volume} {102}},\ \bibinfo {pages} {177202} (\bibinfo {year}
  {2009})}\BibitemShut {NoStop}%
\bibitem [{\citenamefont {Becker}\ \emph {et~al.}(2005)\citenamefont {Becker},
  \citenamefont {Johnsson}, \citenamefont {Kremer},\ and\ \citenamefont
  {Lemmens}}]{Becker2005}%
  \BibitemOpen
  \bibfield  {author} {\bibinfo {author} {\bibfnamefont {R.}~\bibnamefont
  {Becker}}, \bibinfo {author} {\bibfnamefont {M.}~\bibnamefont {Johnsson}},
  \bibinfo {author} {\bibfnamefont {R.~K.}\ \bibnamefont {Kremer}}, \ and\
  \bibinfo {author} {\bibfnamefont {P.}~\bibnamefont {Lemmens}},\ }\bibfield
  {title} {\enquote {\bibinfo {title} {{Crystal structure and magnetic
  properties of Cu$_3$(TeO$_3$)$_2$Br$_2$ -- a layered compound with a new
  Cu(II) coordination polyhedron}},}\ }\href {\doibase
  https://doi.org/10.1016/j.jssc.2005.04.011} {\bibfield  {journal} {\bibinfo
  {journal} {J. Solid State Chem.}\ }\textbf {\bibinfo {volume} {178}},\
  \bibinfo {pages} {2024--2029} (\bibinfo {year} {2005})}\BibitemShut {NoStop}%
\bibitem [{\citenamefont {Zaharko}\ \emph {et~al.}(2008)\citenamefont
  {Zaharko}, \citenamefont {Mesot}, \citenamefont {Salguero}, \citenamefont
  {Valent\'{\i}}, \citenamefont {Zbiri}, \citenamefont {Johnson}, \citenamefont
  {Filinchuk}, \citenamefont {Klemke}, \citenamefont {Kiefer}, \citenamefont
  {Mys'kiv}, \citenamefont {Str\"assle},\ and\ \citenamefont
  {Mutka}}]{Zaharko2008}%
  \BibitemOpen
  \bibfield  {author} {\bibinfo {author} {\bibfnamefont {O.}~\bibnamefont
  {Zaharko}}, \bibinfo {author} {\bibfnamefont {J.}~\bibnamefont {Mesot}},
  \bibinfo {author} {\bibfnamefont {L.~A.}\ \bibnamefont {Salguero}}, \bibinfo
  {author} {\bibfnamefont {R.}~\bibnamefont {Valent\'{\i}}}, \bibinfo {author}
  {\bibfnamefont {M.}~\bibnamefont {Zbiri}}, \bibinfo {author} {\bibfnamefont
  {M.}~\bibnamefont {Johnson}}, \bibinfo {author} {\bibfnamefont
  {Y.}~\bibnamefont {Filinchuk}}, \bibinfo {author} {\bibfnamefont
  {B.}~\bibnamefont {Klemke}}, \bibinfo {author} {\bibfnamefont
  {K.}~\bibnamefont {Kiefer}}, \bibinfo {author} {\bibfnamefont
  {M.}~\bibnamefont {Mys'kiv}}, \bibinfo {author} {\bibfnamefont
  {T.}~\bibnamefont {Str\"assle}}, \ and\ \bibinfo {author} {\bibfnamefont
  {H.}~\bibnamefont {Mutka}},\ }\bibfield  {title} {\enquote {\bibinfo {title}
  {{Tetrahedra system ${\text{Cu}}_{4}{\text{OCl}}_{6}{\text{daca}}_{4}$:
  High-temperature manifold of molecular configurations governing
  low-temperature properties}},}\ }\href {\doibase 10.1103/PhysRevB.77.224408}
  {\bibfield  {journal} {\bibinfo  {journal} {Phys. Rev. B}\ }\textbf {\bibinfo
  {volume} {77}},\ \bibinfo {pages} {224408} (\bibinfo {year}
  {2008})}\BibitemShut {NoStop}%
\bibitem [{\citenamefont {Choi}\ \emph {et~al.}(2014)\citenamefont {Choi},
  \citenamefont {Do}, \citenamefont {Lemmens}, \citenamefont {van Tol},
  \citenamefont {Shin}, \citenamefont {Jeon}, \citenamefont {Skourski},
  \citenamefont {Rhyee},\ and\ \citenamefont {Berger}}]{Choi2014}%
  \BibitemOpen
  \bibfield  {author} {\bibinfo {author} {\bibfnamefont {K.-Y.}\ \bibnamefont
  {Choi}}, \bibinfo {author} {\bibfnamefont {S.}~\bibnamefont {Do}}, \bibinfo
  {author} {\bibfnamefont {P.}~\bibnamefont {Lemmens}}, \bibinfo {author}
  {\bibfnamefont {J.}~\bibnamefont {van Tol}}, \bibinfo {author} {\bibfnamefont
  {J.}~\bibnamefont {Shin}}, \bibinfo {author} {\bibfnamefont {G.~S.}\
  \bibnamefont {Jeon}}, \bibinfo {author} {\bibfnamefont {Y.}~\bibnamefont
  {Skourski}}, \bibinfo {author} {\bibfnamefont {J.-S.}\ \bibnamefont {Rhyee}},
  \ and\ \bibinfo {author} {\bibfnamefont {H.}~\bibnamefont {Berger}},\
  }\bibfield  {title} {\enquote {\bibinfo {title} {{Coexistence of localized
  and collective magnetism in the coupled-spin-tetrahedra system
  ${\mathrm{Cu}}_{4}{\mathrm{Te}}_{5}{\mathrm{O}}_{12}{\mathrm{Cl}}_{4}$}},}\
  }\href {\doibase 10.1103/PhysRevB.90.184402} {\bibfield  {journal} {\bibinfo
  {journal} {Phys. Rev. B}\ }\textbf {\bibinfo {volume} {90}},\ \bibinfo
  {pages} {184402} (\bibinfo {year} {2014})}\BibitemShut {NoStop}%
\bibitem [{\citenamefont {Zhang}\ \emph {et~al.}(2020)\citenamefont {Zhang},
  \citenamefont {Zhao}, \citenamefont {Gautreau}, \citenamefont {Raczkowski},
  \citenamefont {Saha}, \citenamefont {Garlea}, \citenamefont {Cao},
  \citenamefont {Hong}, \citenamefont {Jeschke}, \citenamefont {Mahanti},
  \citenamefont {Birol}, \citenamefont {Assaad},\ and\ \citenamefont
  {Ke}}]{Zhang2020}%
  \BibitemOpen
  \bibfield  {author} {\bibinfo {author} {\bibfnamefont {H.}~\bibnamefont
  {Zhang}}, \bibinfo {author} {\bibfnamefont {Z.}~\bibnamefont {Zhao}},
  \bibinfo {author} {\bibfnamefont {D.}~\bibnamefont {Gautreau}}, \bibinfo
  {author} {\bibfnamefont {M.}~\bibnamefont {Raczkowski}}, \bibinfo {author}
  {\bibfnamefont {A.}~\bibnamefont {Saha}}, \bibinfo {author} {\bibfnamefont
  {V.}~\bibnamefont {Garlea}}, \bibinfo {author} {\bibfnamefont
  {H.}~\bibnamefont {Cao}}, \bibinfo {author} {\bibfnamefont {T.}~\bibnamefont
  {Hong}}, \bibinfo {author} {\bibfnamefont {H.}~\bibnamefont {Jeschke}},
  \bibinfo {author} {\bibfnamefont {S.}~\bibnamefont {Mahanti}}, \bibinfo
  {author} {\bibfnamefont {T.}~\bibnamefont {Birol}}, \bibinfo {author}
  {\bibfnamefont {F.}~\bibnamefont {Assaad}}, \ and\ \bibinfo {author}
  {\bibfnamefont {X.}~\bibnamefont {Ke}},\ }\bibfield  {title} {\enquote
  {\bibinfo {title} {{Coexistence and Interaction of Spinons and Magnons in an
  Antiferromagnet with Alternating Antiferromagnetic and Ferromagnetic Quantum
  Spin Chains}},}\ }\href {\doibase 10.1103/PhysRevLett.125.037204} {\bibfield
  {journal} {\bibinfo  {journal} {Phys. Rev. Lett}\ }\textbf {\bibinfo {volume}
  {125}},\ \bibinfo {pages} {037204} (\bibinfo {year} {2020})}\BibitemShut
  {NoStop}%
\bibitem [{\citenamefont {Effenberger}(1986)}]{kristall113}%
  \BibitemOpen
  \bibfield  {author} {\bibinfo {author} {\bibfnamefont {H.}~\bibnamefont
  {Effenberger}},\ }\bibfield  {title} {\enquote {\bibinfo {title} {{Die
  Kristallstrukturen von drei Modifikationen des Cu(SeO$_3$)}},}\ }\href@noop
  {} {\bibfield  {journal} {\bibinfo  {journal} {Z. Krist}\ }\textbf {\bibinfo
  {volume} {175}},\ \bibinfo {pages} {61--72} (\bibinfo {year}
  {1986})}\BibitemShut {NoStop}%
\bibitem [{\citenamefont {\ifmmode \check{Z}\else
  \v{Z}\fi{}ivkovi\ifmmode~\acute{c}\else \'{c}\fi{}}\ \emph
  {et~al.}(2012)\citenamefont {\ifmmode \check{Z}\else
  \v{Z}\fi{}ivkovi\ifmmode~\acute{c}\else \'{c}\fi{}}, \citenamefont
  {Djoki\ifmmode~\acute{c}\else \'{c}\fi{}}, \citenamefont {Herak},
  \citenamefont {Paji\ifmmode~\acute{c}\else \'{c}\fi{}}, \citenamefont
  {Pr\ifmmode~\check{s}\else \v{s}\fi{}a}, \citenamefont {Pattison},
  \citenamefont {Dominko}, \citenamefont {Mickovi\ifmmode~\acute{c}\else
  \'{c}\fi{}}, \citenamefont {Cin\ifmmode \check{c}\else
  \v{c}\fi{}i\ifmmode~\acute{c}\else \'{c}\fi{}}, \citenamefont {Forr\'o},
  \citenamefont {Berger},\ and\ \citenamefont {R\o{}nnow}}]{SeCuO3_2012_bulk}%
  \BibitemOpen
  \bibfield  {author} {\bibinfo {author} {\bibfnamefont {I.}~\bibnamefont
  {\ifmmode \check{Z}\else \v{Z}\fi{}ivkovi\ifmmode~\acute{c}\else
  \'{c}\fi{}}}, \bibinfo {author} {\bibfnamefont {D.~M.}\ \bibnamefont
  {Djoki\ifmmode~\acute{c}\else \'{c}\fi{}}}, \bibinfo {author} {\bibfnamefont
  {M.}~\bibnamefont {Herak}}, \bibinfo {author} {\bibfnamefont
  {D.}~\bibnamefont {Paji\ifmmode~\acute{c}\else \'{c}\fi{}}}, \bibinfo
  {author} {\bibfnamefont {K.}~\bibnamefont {Pr\ifmmode~\check{s}\else
  \v{s}\fi{}a}}, \bibinfo {author} {\bibfnamefont {P.}~\bibnamefont
  {Pattison}}, \bibinfo {author} {\bibfnamefont {D.}~\bibnamefont {Dominko}},
  \bibinfo {author} {\bibfnamefont {Z.}~\bibnamefont
  {Mickovi\ifmmode~\acute{c}\else \'{c}\fi{}}}, \bibinfo {author}
  {\bibfnamefont {D.}~\bibnamefont {Cin\ifmmode \check{c}\else
  \v{c}\fi{}i\ifmmode~\acute{c}\else \'{c}\fi{}}}, \bibinfo {author}
  {\bibfnamefont {L.}~\bibnamefont {Forr\'o}}, \bibinfo {author} {\bibfnamefont
  {H.}~\bibnamefont {Berger}}, \ and\ \bibinfo {author} {\bibfnamefont {H.~M.}\
  \bibnamefont {R\o{}nnow}},\ }\bibfield  {title} {\enquote {\bibinfo {title}
  {{Site-selective quantum correlations revealed by magnetic anisotropy in the
  tetramer system SeCuO${}_{3}$}},}\ }\href {\doibase
  10.1103/PhysRevB.86.054405} {\bibfield  {journal} {\bibinfo  {journal} {Phys.
  Rev. B}\ }\textbf {\bibinfo {volume} {86}},\ \bibinfo {pages} {054405}
  (\bibinfo {year} {2012})}\BibitemShut {NoStop}%
\bibitem [{\citenamefont {Cvitani\ifmmode~\acute{c}\else \'{c}\fi{}}\ \emph
  {et~al.}(2018)\citenamefont {Cvitani\ifmmode~\acute{c}\else \'{c}\fi{}},
  \citenamefont {\ifmmode~\check{S}\else \v{S}\fi{}urija}, \citenamefont
  {Pr\ifmmode~\check{s}\else \v{s}\fi{}a}, \citenamefont {Zaharko},
  \citenamefont {Kup\ifmmode \check{c}\else \v{c}\fi{}i\ifmmode~\acute{c}\else
  \'{c}\fi{}}, \citenamefont {Babkevich}, \citenamefont {Frontzek},
  \citenamefont {Po\ifmmode~\check{z}\else \v{z}\fi{}ek}, \citenamefont
  {Berger}, \citenamefont {Magrez}, \citenamefont {R\o{}nnow}, \citenamefont
  {Grbi\ifmmode~\acute{c}\else \'{c}\fi{}},\ and\ \citenamefont {\ifmmode
  \check{Z}\else \v{Z}\fi{}ivkovi\ifmmode~\acute{c}\else
  \'{c}\fi{}}}]{SeCuO3_diffraction}%
  \BibitemOpen
  \bibfield  {author} {\bibinfo {author} {\bibfnamefont {T.}~\bibnamefont
  {Cvitani\ifmmode~\acute{c}\else \'{c}\fi{}}}, \bibinfo {author}
  {\bibfnamefont {V.}~\bibnamefont {\ifmmode~\check{S}\else \v{S}\fi{}urija}},
  \bibinfo {author} {\bibfnamefont {K.}~\bibnamefont {Pr\ifmmode~\check{s}\else
  \v{s}\fi{}a}}, \bibinfo {author} {\bibfnamefont {O.}~\bibnamefont {Zaharko}},
  \bibinfo {author} {\bibfnamefont {I.}~\bibnamefont {Kup\ifmmode
  \check{c}\else \v{c}\fi{}i\ifmmode~\acute{c}\else \'{c}\fi{}}}, \bibinfo
  {author} {\bibfnamefont {P.}~\bibnamefont {Babkevich}}, \bibinfo {author}
  {\bibfnamefont {M.}~\bibnamefont {Frontzek}}, \bibinfo {author}
  {\bibfnamefont {M.}~\bibnamefont {Po\ifmmode~\check{z}\else \v{z}\fi{}ek}},
  \bibinfo {author} {\bibfnamefont {H.}~\bibnamefont {Berger}}, \bibinfo
  {author} {\bibfnamefont {A.}~\bibnamefont {Magrez}}, \bibinfo {author}
  {\bibfnamefont {H.~M.}\ \bibnamefont {R\o{}nnow}}, \bibinfo {author}
  {\bibfnamefont {M.~S.}\ \bibnamefont {Grbi\ifmmode~\acute{c}\else
  \'{c}\fi{}}}, \ and\ \bibinfo {author} {\bibfnamefont {I.}~\bibnamefont
  {\ifmmode \check{Z}\else \v{Z}\fi{}ivkovi\ifmmode~\acute{c}\else
  \'{c}\fi{}}},\ }\bibfield  {title} {\enquote {\bibinfo {title} {{Singlet
  state formation and its impact on the magnetic structure in the tetramer
  system ${\mathrm{SeCuO}}_{3}$}},}\ }\href {\doibase
  10.1103/PhysRevB.98.054409} {\bibfield  {journal} {\bibinfo  {journal} {Phys.
  Rev. B}\ }\textbf {\bibinfo {volume} {98}},\ \bibinfo {pages} {054409}
  (\bibinfo {year} {2018})}\BibitemShut {NoStop}%
\bibitem [{\citenamefont {Lee}\ \emph {et~al.}(2017)\citenamefont {Lee},
  \citenamefont {Lee}, \citenamefont {van Tol}, \citenamefont {Kuhns},
  \citenamefont {Reyes}, \citenamefont {Berger},\ and\ \citenamefont
  {Choi}}]{secuo3_lee_dec}%
  \BibitemOpen
  \bibfield  {author} {\bibinfo {author} {\bibfnamefont {S.}~\bibnamefont
  {Lee}}, \bibinfo {author} {\bibfnamefont {W.-J.}\ \bibnamefont {Lee}},
  \bibinfo {author} {\bibfnamefont {J.}~\bibnamefont {van Tol}}, \bibinfo
  {author} {\bibfnamefont {P.~L.}\ \bibnamefont {Kuhns}}, \bibinfo {author}
  {\bibfnamefont {A.~P.}\ \bibnamefont {Reyes}}, \bibinfo {author}
  {\bibfnamefont {H.}~\bibnamefont {Berger}}, \ and\ \bibinfo {author}
  {\bibfnamefont {K.-Y.}\ \bibnamefont {Choi}},\ }\bibfield  {title} {\enquote
  {\bibinfo {title} {{Anomalous spin dynamics in the coupled spin tetramer
  system ${\mathrm{CuSeO}}_{3}$}},}\ }\href {\doibase
  10.1103/PhysRevB.95.054405} {\bibfield  {journal} {\bibinfo  {journal} {Phys.
  Rev. B}\ }\textbf {\bibinfo {volume} {95}},\ \bibinfo {pages} {054405}
  (\bibinfo {year} {2017})}\BibitemShut {NoStop}%
\bibitem [{\citenamefont {Novosel}\ \emph {et~al.}(2019)\citenamefont
  {Novosel}, \citenamefont {Lafargue-Dit-Hauret}, \citenamefont
  {Rapljenovi\'{c}}, \citenamefont {Dragi\ifmmode \check{c}\else
  \v{c}\fi{}evi\ifmmode~\acute{c}\else \'{c}\fi{}}, \citenamefont {Berger},
  \citenamefont {Cin\ifmmode \check{c}\else \v{c}\fi{}i\ifmmode~\acute{c}\else
  \'{c}\fi{}}, \citenamefont {Rocquefelte},\ and\ \citenamefont
  {Herak}}]{SeCuO3_decoupled}%
  \BibitemOpen
  \bibfield  {author} {\bibinfo {author} {\bibfnamefont {N.}~\bibnamefont
  {Novosel}}, \bibinfo {author} {\bibfnamefont {W.}~\bibnamefont
  {Lafargue-Dit-Hauret}}, \bibinfo {author} {\bibfnamefont {Z.}~\bibnamefont
  {Rapljenovi\'{c}}}, \bibinfo {author} {\bibfnamefont {M.}~\bibnamefont
  {Dragi\ifmmode \check{c}\else \v{c}\fi{}evi\ifmmode~\acute{c}\else
  \'{c}\fi{}}}, \bibinfo {author} {\bibfnamefont {H.}~\bibnamefont {Berger}},
  \bibinfo {author} {\bibfnamefont {D.}~\bibnamefont {Cin\ifmmode
  \check{c}\else \v{c}\fi{}i\ifmmode~\acute{c}\else \'{c}\fi{}}}, \bibinfo
  {author} {\bibfnamefont {X.}~\bibnamefont {Rocquefelte}}, \ and\ \bibinfo
  {author} {\bibfnamefont {M.}~\bibnamefont {Herak}},\ }\bibfield  {title}
  {\enquote {\bibinfo {title} {{Strong decoupling between magnetic subsystems
  in the low-dimensional spin-$\frac{1}{2}$ antiferromagnet
  ${\text{SeCuO}}_{3}$}},}\ }\href {\doibase 10.1103/PhysRevB.99.014434}
  {\bibfield  {journal} {\bibinfo  {journal} {Phys. Rev. B}\ }\textbf {\bibinfo
  {volume} {99}},\ \bibinfo {pages} {014434} (\bibinfo {year}
  {2019})}\BibitemShut {NoStop}%
\bibitem [{\citenamefont {\v{S}urija}\ \emph {et~al.}(2013)\citenamefont
  {\v{S}urija}, \citenamefont {Pr\v{s}a}, \citenamefont {R\o{}nnow},
  \citenamefont {Boehm},\ and\ \citenamefont {\v{Z}ivkovi\'c}}]{DOI_IN8}%
  \BibitemOpen
  \bibfield  {author} {\bibinfo {author} {\bibfnamefont {V.}~\bibnamefont
  {\v{S}urija}}, \bibinfo {author} {\bibfnamefont {K.}~\bibnamefont
  {Pr\v{s}a}}, \bibinfo {author} {\bibfnamefont {H.~M.}\ \bibnamefont
  {R\o{}nnow}}, \bibinfo {author} {\bibfnamefont {M.}~\bibnamefont {Boehm}}, \
  and\ \bibinfo {author} {\bibfnamefont {I.}~\bibnamefont {\v{Z}ivkovi\'c}},\
  }\bibfield  {title} {\enquote {\bibinfo {title} {{SeCuO$_3$: a coupled
  cluster system close to criticality}},}\ }\href {\doibase
  doi:10.5291/ILL-DATA.4-01-1295} {\  (\bibinfo {year} {2013}),\
  doi:10.5291/ILL-DATA.4-01-1295}\BibitemShut {NoStop}%
\bibitem [{\citenamefont {\v{S}urija}\ \emph {et~al.}(2015)\citenamefont
  {\v{S}urija}, \citenamefont {Pr\v{s}a}, \citenamefont {R\o{}nnow},
  \citenamefont {Steffens},\ and\ \citenamefont {\v{Z}ivkovi\'c}}]{DOI_thales}%
  \BibitemOpen
  \bibfield  {author} {\bibinfo {author} {\bibfnamefont {V.}~\bibnamefont
  {\v{S}urija}}, \bibinfo {author} {\bibfnamefont {K.}~\bibnamefont
  {Pr\v{s}a}}, \bibinfo {author} {\bibfnamefont {H.~M.}\ \bibnamefont
  {R\o{}nnow}}, \bibinfo {author} {\bibfnamefont {P.}~\bibnamefont {Steffens}},
  \ and\ \bibinfo {author} {\bibfnamefont {I.}~\bibnamefont {\v{Z}ivkovi\'c}},\
  }\bibfield  {title} {\enquote {\bibinfo {title} {{Low-energy magnetic modes
  in SeCuO$_3$}},}\ }\href {\doibase doi:10.5291/ILL-DATA.4-01-1459} {\
  (\bibinfo {year} {2015}),\ doi:10.5291/ILL-DATA.4-01-1459}\BibitemShut
  {NoStop}%
\bibitem [{sm()}]{sm}%
  \BibitemOpen
  \href@noop {} {\bibinfo  {journal} {{Details are provided in the Supplemental
  Material at the end of this document}}\ }\BibitemShut {NoStop}%
\bibitem [{\citenamefont {Toth}\ and\ \citenamefont {Lake}(2015)}]{spinw}%
  \BibitemOpen
\bibfield  {journal} {  }\bibfield  {author} {\bibinfo {author} {\bibfnamefont
  {S.}~\bibnamefont {Toth}}\ and\ \bibinfo {author} {\bibfnamefont
  {B.}~\bibnamefont {Lake}},\ }\bibfield  {title} {\enquote {\bibinfo {title}
  {{Linear spin wave theory for single-Q incommensurate magnetic
  structures}},}\ }\href {\doibase 10.1088/0953-8984/27/16/166002} {\bibfield
  {journal} {\bibinfo  {journal} {J. Phys. Condens. Matter}\ }\textbf {\bibinfo
  {volume} {27}},\ \bibinfo {pages} {166002} (\bibinfo {year}
  {2015})}\BibitemShut {NoStop}%
\bibitem [{\citenamefont {Cowley}\ \emph {et~al.}(1969)\citenamefont {Cowley},
  \citenamefont {Buyers}, \citenamefont {Martel},\ and\ \citenamefont
  {Stevenson}}]{Cowley1969}%
  \BibitemOpen
  \bibfield  {author} {\bibinfo {author} {\bibfnamefont {R.~A.}\ \bibnamefont
  {Cowley}}, \bibinfo {author} {\bibfnamefont {W.~J.~L.}\ \bibnamefont
  {Buyers}}, \bibinfo {author} {\bibfnamefont {P.}~\bibnamefont {Martel}}, \
  and\ \bibinfo {author} {\bibfnamefont {R.~W.~H.}\ \bibnamefont {Stevenson}},\
  }\bibfield  {title} {\enquote {\bibinfo {title} {{Two-Magnon Scattering of
  Neutrons}},}\ }\href {\doibase 10.1103/PhysRevLett.23.86} {\bibfield
  {journal} {\bibinfo  {journal} {Phys. Rev. Lett.}\ }\textbf {\bibinfo
  {volume} {23}},\ \bibinfo {pages} {86--89} (\bibinfo {year}
  {1969})}\BibitemShut {NoStop}%
\bibitem [{\citenamefont {Huberman}\ \emph {et~al.}(2005)\citenamefont
  {Huberman}, \citenamefont {Coldea}, \citenamefont {Cowley}, \citenamefont
  {Tennant}, \citenamefont {Leheny}, \citenamefont {Christianson},\ and\
  \citenamefont {Frost}}]{Huberman2mag}%
  \BibitemOpen
  \bibfield  {author} {\bibinfo {author} {\bibfnamefont {T.}~\bibnamefont
  {Huberman}}, \bibinfo {author} {\bibfnamefont {R.}~\bibnamefont {Coldea}},
  \bibinfo {author} {\bibfnamefont {R.~A.}\ \bibnamefont {Cowley}}, \bibinfo
  {author} {\bibfnamefont {D.~A.}\ \bibnamefont {Tennant}}, \bibinfo {author}
  {\bibfnamefont {R.~L.}\ \bibnamefont {Leheny}}, \bibinfo {author}
  {\bibfnamefont {R.~J.}\ \bibnamefont {Christianson}}, \ and\ \bibinfo
  {author} {\bibfnamefont {C.~D.}\ \bibnamefont {Frost}},\ }\bibfield  {title}
  {\enquote {\bibinfo {title} {{Two-magnon excitations observed by neutron
  scattering in the two-dimensional spin-$\frac{5}{2}$ Heisenberg
  antiferromagnet ${\mathrm{Rb}}_{2}\mathrm{Mn}{\mathrm{F}}_{4}$}},}\ }\href
  {\doibase 10.1103/PhysRevB.72.014413} {\bibfield  {journal} {\bibinfo
  {journal} {Phys. Rev. B}\ }\textbf {\bibinfo {volume} {72}},\ \bibinfo
  {pages} {014413} (\bibinfo {year} {2005})}\BibitemShut {NoStop}%
\bibitem [{\citenamefont {Mezio}\ \emph {et~al.}(2011)\citenamefont {Mezio},
  \citenamefont {Sposetti}, \citenamefont {Manuel},\ and\ \citenamefont
  {Trumper}}]{Mezio2011}%
  \BibitemOpen
  \bibfield  {author} {\bibinfo {author} {\bibfnamefont {A.}~\bibnamefont
  {Mezio}}, \bibinfo {author} {\bibfnamefont {C.~N.}\ \bibnamefont {Sposetti}},
  \bibinfo {author} {\bibfnamefont {L.~O.}\ \bibnamefont {Manuel}}, \ and\
  \bibinfo {author} {\bibfnamefont {A.~E.}\ \bibnamefont {Trumper}},\
  }\bibfield  {title} {\enquote {\bibinfo {title} {{A test of the bosonic
  spinon theory for the triangular antiferromagnet spectrum}},}\ }\href
  {\doibase 10.1209/0295-5075/94/47001} {\bibfield  {journal} {\bibinfo
  {journal} {Europhys. Lett.}\ }\textbf {\bibinfo {volume} {94}},\ \bibinfo
  {pages} {47001} (\bibinfo {year} {2011})}\BibitemShut {NoStop}%
\bibitem [{\citenamefont {Shao}\ \emph {et~al.}(2017)\citenamefont {Shao},
  \citenamefont {Qin}, \citenamefont {Capponi}, \citenamefont {Chesi},
  \citenamefont {Meng},\ and\ \citenamefont {Sandvik}}]{Shao2017}%
  \BibitemOpen
  \bibfield  {author} {\bibinfo {author} {\bibfnamefont {H.}~\bibnamefont
  {Shao}}, \bibinfo {author} {\bibfnamefont {Y.~Q.}\ \bibnamefont {Qin}},
  \bibinfo {author} {\bibfnamefont {S.}~\bibnamefont {Capponi}}, \bibinfo
  {author} {\bibfnamefont {S.}~\bibnamefont {Chesi}}, \bibinfo {author}
  {\bibfnamefont {Z.~Y.}\ \bibnamefont {Meng}}, \ and\ \bibinfo {author}
  {\bibfnamefont {A.~W.}\ \bibnamefont {Sandvik}},\ }\bibfield  {title}
  {\enquote {\bibinfo {title} {{Nearly Deconfined Spinon Excitations in the
  Square-Lattice Spin-$1/2$ Heisenberg Antiferromagnet}},}\ }\href {\doibase
  10.1103/PhysRevX.7.041072} {\bibfield  {journal} {\bibinfo  {journal} {Phys.
  Rev. X}\ }\textbf {\bibinfo {volume} {7}},\ \bibinfo {pages} {041072}
  (\bibinfo {year} {2017})}\BibitemShut {NoStop}%
\bibitem [{\citenamefont {Yu}\ \emph {et~al.}(2018)\citenamefont {Yu},
  \citenamefont {Wang}, \citenamefont {Dong}, \citenamefont {Yao},\ and\
  \citenamefont {Li}}]{Yu2018}%
  \BibitemOpen
  \bibfield  {author} {\bibinfo {author} {\bibfnamefont {S.-L.}\ \bibnamefont
  {Yu}}, \bibinfo {author} {\bibfnamefont {W.}~\bibnamefont {Wang}}, \bibinfo
  {author} {\bibfnamefont {Z.-Y.}\ \bibnamefont {Dong}}, \bibinfo {author}
  {\bibfnamefont {Z.-J.}\ \bibnamefont {Yao}}, \ and\ \bibinfo {author}
  {\bibfnamefont {J.-X.}\ \bibnamefont {Li}},\ }\bibfield  {title} {\enquote
  {\bibinfo {title} {{Deconfinement of spinons in frustrated spin systems:
  Spectral perspective}},}\ }\href {\doibase 10.1103/PhysRevB.98.134410}
  {\bibfield  {journal} {\bibinfo  {journal} {Phys. Rev. B}\ }\textbf {\bibinfo
  {volume} {98}},\ \bibinfo {pages} {134410} (\bibinfo {year}
  {2018})}\BibitemShut {NoStop}%
\bibitem [{\citenamefont {Ghioldi}\ \emph {et~al.}(2018)\citenamefont
  {Ghioldi}, \citenamefont {Gonzalez}, \citenamefont {Zhang}, \citenamefont
  {Kamiya}, \citenamefont {Manuel}, \citenamefont {Trumper},\ and\
  \citenamefont {Batista}}]{Ghioldi2018}%
  \BibitemOpen
  \bibfield  {author} {\bibinfo {author} {\bibfnamefont {E.~A.}\ \bibnamefont
  {Ghioldi}}, \bibinfo {author} {\bibfnamefont {M.~G.}\ \bibnamefont
  {Gonzalez}}, \bibinfo {author} {\bibfnamefont {S.-S.}\ \bibnamefont {Zhang}},
  \bibinfo {author} {\bibfnamefont {Y.}~\bibnamefont {Kamiya}}, \bibinfo
  {author} {\bibfnamefont {L.~O.}\ \bibnamefont {Manuel}}, \bibinfo {author}
  {\bibfnamefont {A.~E.}\ \bibnamefont {Trumper}}, \ and\ \bibinfo {author}
  {\bibfnamefont {C.~D.}\ \bibnamefont {Batista}},\ }\bibfield  {title}
  {\enquote {\bibinfo {title} {{Dynamical structure factor of the triangular
  antiferromagnet: Schwinger boson theory beyond mean field}},}\ }\href
  {\doibase 10.1103/PhysRevB.98.184403} {\bibfield  {journal} {\bibinfo
  {journal} {Phys. Rev. B}\ }\textbf {\bibinfo {volume} {98}},\ \bibinfo
  {pages} {184403} (\bibinfo {year} {2018})}\BibitemShut {NoStop}%
\bibitem [{\citenamefont {Ferrari}\ and\ \citenamefont
  {Becca}(2019)}]{Ferrari2019}%
  \BibitemOpen
  \bibfield  {author} {\bibinfo {author} {\bibfnamefont {F.}~\bibnamefont
  {Ferrari}}\ and\ \bibinfo {author} {\bibfnamefont {F.}~\bibnamefont
  {Becca}},\ }\bibfield  {title} {\enquote {\bibinfo {title} {{Dynamical
  Structure Factor of the $J_1$-$J_2$ Heisenberg Model on the Triangular
  Lattice: Magnons, Spinons, and Gauge Fields}},}\ }\href {\doibase
  10.1103/PhysRevX.9.031026} {\bibfield  {journal} {\bibinfo  {journal} {Phys.
  Rev. X}\ }\textbf {\bibinfo {volume} {9}},\ \bibinfo {pages} {031026}
  (\bibinfo {year} {2019})}\BibitemShut {NoStop}%
\end{thebibliography}%

\renewcommand*{\citenumfont}[1]{S#1}
\renewcommand*{\bibnumfmt}[1]{[S#1]}
\setcounter{figure}{0}
\renewcommand{\thefigure}{S\arabic{figure}}
\setcounter{equation}{0}
\renewcommand{\theequation}{S\arabic{equation}}
\setcounter{table}{0}
\renewcommand{\thetable}{S\arabic{table}}

\newpage
\vspace{30cm}
\newpage

\onecolumngrid

\centerline{\large {\bf {Supplemental Material for ``Triplons, Magnons, 
and Spinons in a}}}
 
\vskip1mm

\centerline{\large {\bf {Single Quantum Spin System: SeCuO$_3$''}}}

\vskip4mm

\centerline{L. Testa, V. \v{S}urija, K. Pr\v{s}a, P. Steffens, 
M. Boehm, P. Bourges,}

\centerline{H. Berger, B. Normand, H. M. R\o nnow, and I. \v{Z}ivkovi\'c}

\vskip8mm

\subsection{S1. Experimental Details}

The sample was mounted on an Al holder, using Laue x-ray backscattering 
to orient it in the (\textit{hkh}) scattering plane. Thermal neutron 
measurements on IN8 used an incident neutron wave vector of 2.66 \AA$^{-1}$ 
and the energy resolution at $\omega = 27$ meV was $1.8(2)$ meV (FWHM). In 
the cold neutron measurements on ThALES and 4F1, the incident neutron wave 
vector was 1.55 \AA$^{-1}$ and the resolution at $\omega = 5$ meV was $0.19(5)$ 
meV in both cases; both experiments used a PG(002) monochromator and analyzer, 
and a Be filter placed after the sample to remove higher-order scattering 
processes. The crystal was reoriented in the (\textit{hk$\bar{h}$}) scattering 
plane for the 4F1 experiment. Counting times in the ThALES experiment were 5 
minutes per ${\bf q}$-point and on 4F1 3 minutes per point (Figs.~\ref{lowE_cuts} and \ref{swfitandintensities} of 
the main text). On IN8, each ${\bf q}$-point at temperatures below and directly 
above the ordering transition was measured for 4 minutes [Figs.~\ref{highE}(a-c)],
whereas one measurement during the studies of temperature-dependence lasted 
for 30 s [Figs.~\ref{highE}(d-f)]. The measured intensities, $I({\bf q},\omega)$, can 
be regarded as $S({\bf q},\omega)$ integrated over the resolution functions 
of each instrument. 

\subsection{S2. Four-site model}
\label{sfsm}

Each of the two different four-spin systems represented in Fig.~\ref{fig:structure}(e) of the 
main text has the Heisenberg Hamiltonian 
\begin{equation}
\ham_t = J_{12}^\gamma \big( \Sh_{1_2} \! \cdot \! \Sh_{2_1} + \Sh_{3_1} \! \cdot \! 
\Sh_{4_2} \big) + J_{D} \, \Sh_{2_1} \! \cdot \! \Sh_{3_1}. 
\end{equation}
The ground-state and lowest-lying excited energies obtained by diagonalization 
in the 16$\times$16 Hilbert space are 
\begin{equation}
E_0 = {\textstyle \frac{1}{4}} \Big( -J_D - 2J_{12}^\gamma - 2\sqrt{J_D^2 - 2J_{D} 
J_{12}^\gamma + 4 (J_{12}^\gamma)^2} \Big), \;\;\;\;\;\;\;\;
E_1 = {\textstyle \frac{1}{4}} \Big( -J_{D} - 2 \sqrt{J_{D}^2 + (J_{12}^\gamma)^2} 
\Big),
\label{ee}
\end{equation}
with corresponding eigenstates 
\begin{equation}
\begin{alignedat}{2}
& \ket{\Phi_0}  && = \ket{\up\up\down\down} - A \ket{\up\down\up\down} + B 
\ket{\up\down\down\up} + B \ket{\down\up\up\down} - A \ket{\down\up\down\up}
 + \ket{\down\down\up\up}, \\
& \ket{\Phi_1^-} && = - \ket{\up\down\down\down} + C \ket{\down\up\down\down}
 - C \ket{\down\down\up\down} + \ket{\down\down\down\up}, \\
& \ket{\Phi_1^0} && = - \ket{\up\up\down\down} + D \ket{\up\down\up\down}
 - D \ket{\down\up\down\up} + \ket{\down\down\up\up}, \\
& \ket{\Phi_1^+} && = - \ket{\up\up\up\down} + C \ket{\up\up\down\up} - C 
\ket{\up\down\up\up} + \ket{\down\up\up\up},
\end{alignedat}
\end{equation}
in which the coefficients are given by 
\begin{equation}
\begin{alignedat}{2}
& A = \frac{2J_{12}^\gamma + \sqrt{J_{D}^2 - 2J_{D} J_{12}^\gamma + 4 (J_{12}^\gamma)^2}}
{J_{D}} , \;\;\;\;\;\;\;\;\;\;\;\;\;\;\;\; && C = \frac{J_{D} + \sqrt{J_{D}^2 + 
(J_{12}^\gamma)^2}}{J_{12}} , \\
& B = \frac{2J_{12}^\gamma \big( 2J_{12}^\gamma +\sqrt{J_{D}^2 - 2J_{D} J_{12}^\gamma
 + 4(J_{12}^\gamma)^2} \big)}{J_{D} \big( J_{D} + \sqrt{J_{D}^2 - 2J_{D} J_{12}^\gamma 
 + 4(J_{12}^\gamma)^2} \big)}, && D = \frac{J_{12}^\gamma + \sqrt{J_{D}^2 + 
(J_{12}^\gamma)^2}}{J_{D}}.
\end{alignedat}
\end{equation}
In the limit of strong coupling on the Cu$_1$ dimer ($J_{D} \gg J_{12}^\gamma$), 
it is easy to see that $A, D \rightarrow 1$, $B \rightarrow 0$, and $C \gg 1$. 
The eigenstates may then be reexpressed as 
\begin{equation}
\label{es}
\begin{alignedat}{2}
& \ket{\Phi_0}  && = \ket{\up\up\down\down} - \ket{\up\down\up\down}
 + \ket{\down\down\up\up} - \ket{\down\up\down\up} \;\;\;\;\;\;\;
 = \; \ket{s_1} \otimes \ket{s_2}, \\
& \ket{\Phi_1^-}  && = \ket{\down\down\down\up} - \ket{\up\down\down\down}
 + C \left(\ket{\down\up\down\down} - \ket{\down\down\up\down} \right) \;
 = \; C \ket{s_1} \otimes \ket{t_2^-} - \ket{s_2} \otimes \ket{t_1^-}, \\
& \ket{\Phi_1^0}  && = \ket{\up\down\up\down} - \ket{\up\up\down\down}
 + \ket{\down\down\up\up} - \ket{\down\up\down\up} \;\;\;\;\;\;\;
 = \; - \ket{s_1} \otimes \ket{t_2^0}, \\
& \ket{\Phi_1^+}  && = \ket{\down\up\up\up} - \ket{\up\up\up\down} 
 + C \left(\ket{\up\up\down\up} - \ket{\up\down\up\up} \right) \;
 = \; C \ket{s_1} \otimes \ket{t_2^+} - \ket{t_1^+} \otimes \ket{s_2}, 
\end{alignedat}
\end{equation}
where on the right side we have introduced a singlet-triplet notation.
Thus we use the energy difference, $E_1 - E_0$ in Eq.~(\ref{ee}), to deduce 
Eq.~(\ref{eqt_tetramer_jeff}) of the main text and we use Eq.~(\ref{es}) to demonstrate that, as 
a singlet-triplet energy separation on the two Cu$_2$ sites, it corresponds 
in the limit of strong $J_D$ to an effective magnetic interaction between 
these sites.  

\end{document}